\documentstyle[12pt]{article}

\def\hybrid{\topmargin -20pt    \oddsidemargin 0pt
        \headheight 0pt \headsep 0pt
        \textwidth 6.25in       
        \textheight 9.5in       
        \marginparwidth .875in
        \parskip 5pt plus 1pt   \jot = 1.5ex}

\hybrid

\def\baselinestretch{1.2}

\catcode`\@=11

\def\marginnote#1{}
\newcount\hour
\newcount\minute
\newtoks\amorpm
\hour=\time\divide\hour by60
\minute=\time{\multiply\hour by60 \global\advance\minute by-\hour}
\edef\standardtime{{\ifnum\hour<12 \global\amorpm={am}%
        \else\global\amorpm={pm}\advance\hour by-12 \fi
        \ifnum\hour=0 \hour=12 \fi
        \number\hour:\ifnum\minute<10 0\fi\number\minute\the\amorpm}}
\edef\militarytime{\number\hour:\ifnum\minute<10 0\fi\number\minute}

\def\draftlabel#1{{\@bsphack\if@filesw {\let\thepage\relax
   \xdef\@gtempa{\write\@auxout{\string
      \newlabel{#1}{{\@currentlabel}{\thepage}}}}}\@gtempa
   \if@nobreak \ifvmode\nobreak\fi\fi\fi\@esphack}
        \gdef\@eqnlabel{#1}}
\def\@eqnlabel{}
\def\@vacuum{}
\def\draftmarginnote#1{\marginpar{\raggedright\scriptsize\tt#1}}

\def\draft{\oddsidemargin -.5truein
        \def\@oddfoot{\sl preliminary draft \hfil
        \rm\thepage\hfil\sl\today\quad\militarytime}
        \let\@evenfoot\@oddfoot \overfullrule 3pt
        \let\label=\draftlabel
        \let\marginnote=\draftmarginnote
   \def\@eqnnum{(\theequation)\rlap{\kern\marginparsep\tt\@eqnlabel}%
\global\let\@eqnlabel\@vacuum}  }

\def\preprint{\twocolumn\sloppy\flushbottom\parindent 2em
        \leftmargini 2em\leftmarginv .5em\leftmarginvi .5em
        \oddsidemargin -.5in    \evensidemargin -.5in
        \columnsep .4in \footheight 0pt
        \textwidth 10.in        \topmargin  -.4in
        \headheight 12pt \topskip .4in
        \textheight 6.9in \footskip 0pt
        \def\@oddhead{\thepage\hfil\addtocounter{page}{1}\thepage}
        \let\@evenhead\@oddhead \def\@oddfoot{} \def\@evenfoot{} }

\def\numberbysection{\@addtoreset{equation}{section}
        \def\theequation{\thesection.\arabic{equation}}}

\def\underline#1{\relax\ifmmode\@@underline#1\else
        $\@@underline{\hbox{#1}}$\relax\fi}

\def\titlepage{\@restonecolfalse\if@twocolumn\@restonecoltrue\onecolumn
     \else \newpage \fi \thispagestyle{empty}\c@page\z@
        \def\thefootnote{\fnsymbol{footnote}} }

\def\endtitlepage{\if@restonecol\twocolumn \else \newpage \fi
        \def\thefootnote{\arabic{footnote}}
        \setcounter{footnote}{0}}  

\catcode`@=12
\relax

\def\figcap{\section*{Figure Captions\markboth
        {FIGURECAPTIONS}{FIGURECAPTIONS}}\list
        {Figure \arabic{enumi}:\hfill}{\settowidth\labelwidth{Figure
999:}
        \leftmargin\labelwidth
        \advance\leftmargin\labelsep\usecounter{enumi}}}
 \relax
\def\tablecap{\section*{Table Captions\markboth
        {TABLECAPTIONS}{TABLECAPTIONS}}\list
        {Table \arabic{enumi}:\hfill}{\settowidth\labelwidth{Table
999:}
        \leftmargin\labelwidth
        \advance\leftmargin\labelsep\usecounter{enumi}}}
 \relax
\def\reflist{\section*{References\markboth
        {REFLIST}{REFLIST}}\list
        {[\arabic{enumi}]\hfill}{\settowidth\labelwidth{[999]}
        \leftmargin\labelwidth
        \advance\leftmargin\labelsep\usecounter{enumi}}}
 \relax
\makeatletter
\newcounter{pubctr}
\def\publist{\@ifnextchar[{\@publist}{\@@publist}}
\def\@publist[#1]{\list
        {[\arabic{pubctr}]\hfill}{\settowidth\labelwidth{[999]}
        \leftmargin\labelwidth
        \advance\leftmargin\labelsep
        \@nmbrlisttrue\def\@listctr{pubctr}
        \setcounter{pubctr}{#1}\addtocounter{pubctr}{-1}}}
\def\@@publist{\list
        {[\arabic{pubctr}]\hfill}{\settowidth\labelwidth{[999]}
        \leftmargin\labelwidth
        \advance\leftmargin\labelsep
        \@nmbrlisttrue\def\@listctr{pubctr}}}
 \relax
\makeatother
\newskip\humongous \humongous=0pt plus 1000pt minus 1000pt

\newif\ifdtup

\relax

\def\be{\begin{equation}}
\def\ee{\end{equation}}
\def\ba{\begin{eqnarray}}
\def\ea{\end{eqnarray}}

\def\del{\partial}

\def\r{\rho}
\def\a{\alpha}

\def\b{\beta}

\def\d{\delta}
\def\D{\Delta}

\def\th{\theta}

\def\m{\mu}
\def\n{\nu}
\def\om{\omega}
\def\Om{\Omega}
\def\l{\lambda}

\def\s{\sigma}

\def\no{\noindent}

\def\qq{\qquad}

\def\IR{\relax{\rm I\kern-.18em R}}

\def \ha {{1\over 2}}

\def \ov {\over}

\def\IR{\relax{\rm I\kern-.18em R}}
\def\inv{^{\raise.15ex\hbox{${\scriptscriptstyle -}$}\kern-.05em 1}}

\def\tL{{\tilde L}}

\begin{document}

\renewcommand{\theequation}{\thesection.\arabic{equation}}

\newcommand{\beq}{\begin{equation}}
\newcommand{\eeq}[1]{\label{#1}\end{equation}}
\newcommand{\ber}{\begin{eqnarray}}
\newcommand{\eer}[1]{\label{#1}\end{eqnarray}}
\newcommand{\eqn}[1]{(\ref{#1})}
\begin{titlepage}
\begin{center}

\hfill CERN-TH/99-270\\
\hfill hep--th/9909041\\

\vskip .6in



{\large \bf States and Curves of Five-Dimensional Gauged Supergravity}

\vskip 0.5in

{\bf Ioannis Bakas${}^1$ }\phantom{x}and\phantom{x} 
{\bf Konstadinos Sfetsos${}^2$}
\vskip 0.1in
{\em ${}^1\!$Department of Physics, University of Patras \\
GR-26500 Patras, Greece\\
{\tt bakas@nxth04.cern.ch, ajax.physics.upatras.gr}}\\
\vskip .2in
{\em ${}^2\!$Theory Division, CERN\\
     CH-1211 Geneva 23, Switzerland\\
{\tt sfetsos@mail.cern.ch}}\\

\end{center}

\vskip .5in

\centerline{\bf Abstract}

\no
We consider the sector of ${\cal N}=8$ five-dimensional gauged 
supergravity with non-trivial scalar fields in the coset space
$SL(6,\IR)/SO(6)$, plus the metric. We find that the most general
supersymmetric solution is parametrized by six 
real moduli and analyze its properties using the theory of 
algebraic curves. In the generic case, where no continuous subgroup 
of the original $SO(6)$ symmetry remains unbroken, the algebraic 
curve of the corresponding solution is a Riemann surface of genus seven. 
When some cycles shrink to zero size the symmetry group is enhanced, 
whereas the genus of the Riemann surface is lowered accordingly.
The uniformization of the curves is carried out explicitly and yields 
various supersymmetric configurations in terms of elliptic functions.
We also analyze the ten-dimensional type-IIB supergravity origin of 
our solutions and show that they represent the gravitational field of 
a large number of D3-branes continuously 
distributed on hyper-surfaces embedded 
in the six-dimensional space transverse to the branes.
The spectra of massless scalar and graviton excitations
are also studied on these backgrounds by casting the associated 
differential equations into Schr\"odinger equations with non-trivial
potentials. The potentials are found to be of Calogero type, rational
or elliptic, depending on the background configuration that is used.

\vskip .5cm
\noindent
CERN-TH/99-270\\
September 1999
\end{titlepage}
\vfill
\eject

\def\baselinestretch{1.2}
\baselineskip 16 pt
\noindent

\def\tT{{\tilde T}}
\def\tg{{\tilde g}}
\def\tL{{\tilde L}}


\section{Introduction}

Ungauged and gauged ${\cal N}=8$ supergravities in five dimensions were 
constructed several years ago in \cite{cremmer} and \cite{PPN,GRW}, 
following the analogous construction made in four 
dimensions in \cite{CJ} and \cite{WN1}. More recently
it has become clear that solutions of five-dimensional gauged
supergravity play an important r\^ole in the context of the AdS/CFT
correspondence \cite{Maldacena,Witten,GKP}.
In particular the maximum supersymmetric vacuum state in five-dimensional
gauged supergravity with $AdS_5$ 
geometry, originates from the $AdS_5\times S^5$ solution in ten-dimensional
type-IIB supergravity. The latter solution arises as the near horizon 
geometry of the solution representing the gravitational field of a large 
number of coincident D3-branes and has been conjectured to provide
the correct 
framework for analyzing ${\cal N}=4$ supersymmetric $SU(N)$ Yang--Mills for 
large $N$ and 't Hooft coupling constant at the conformal point of the Coulomb
branch. 

The supergravity approach to gauge theories at strong coupling is
applicable not only at conformality, but also away from it. 
In particular, when the six scalar fields of the ${\cal N}=4$ supersymmetric
Yang--Mills theory
acquire Higgs expectation values we move away from the origin 
of the Coulomb branch and the appropriate supergravity solution corresponds
to a multicenter distribution of D3-branes with the centers, where the branes 
are located, associated with the
scalar Higgs expectation values in the gauge theory side. 
A prototype example of such D3-brane distributions is the two-center solution
that has been studied in \cite{Maldacena,MW,KW}, whereas
examples of continuous D3-brane distributions arise naturally 
in the supersymmetric limit of rotating D3-brane solutions \cite{KLT,sfe1}.
Concentrating on the case of continuous distributions, note 
that from a ten-dimensional type-IIB supergravity view point the 
$SO(6)$ symmetry, associated with the round $S^5$-sphere, is broken
because this sphere is deformed. On the other hand,
from the point of view of five-dimensional gauged supergravity 
the deformation of the sphere is associated with the fact that
some of the scalar fields in the theory are turned on. 
Hence, finding solutions of 
five-dimensional gauged supersgravity might shed more light into the AdS/CFT
correspondence as far as the Coulomb branch is concerned. 
Using such solutions, investigations 
of the spectrum of massless scalars excitations 
and of the quark-antiquark potential have already been carried out
with sometimes suprising results \cite{FGPW2,BS1,CR-GR}. 
Solutions of the five-dimensional theory 
are also important in a non-perturbative
treatment of the renormalization group flow in gauge theories 
at strong coupling \cite{GPPZ1,DZ,KPW,FGPW1}.

An additional motivation for studying solutions of five-dimensional gauged
supergravity is the fact that for a class of such configurations, 
four-dimensional
Poincar\'e invariance is preserved. It turns out that our four-dimensional 
space-time can be viewed as being embedded
non-trivially in the five-dimensional solution
with a warp factor. This particular idea of our space-time 
as a membrane in higher dimensions is quite old \cite{GW} 
and has been recently 
revived with interesting phenomenological consequences on the mass
hierarchy problem \cite{rasu}. In that work, in particular, our 
four-dimensional world was embedded into the 
$AdS_5$ space from which a slice was cut out; it results into 
a normalizable graviton zero mode, but also to a continuum spectrum
of massive ones above it with no mass gap separating them.
The use of more general solutions of five-dimensional gauged supergravity
certainly creates more possibilities and in fact there are solutions with
a mass gap that separates the massless mode from the massive ones \cite{BS2}.

This paper is organized as follows: In section 2 we present a brief summary 
of some basic facts about
${\cal N}=8$ five-dimensional gauged supergavity with gauge
group $SO(6)$. In particular, we restrict our attention to the sector of
the theory where only the
metric and the scalar fields associated with the coset space $SL(6,\IR)/SO(6)$ 
are turned one. In section 3 we find the most general supersymmetric 
configuration in this sector, which as it turns out, depends on six 
real moduli. Our solutions have a ten-dimensional 
origin within type-IIB supergravity and represent the 
gravitational field of continuous distributions of D3-branes in hyper-surfaces
embedded in the transverse space to the branes.
In section 4 we further analyze our solution using some concepts from the 
theory of algebraic curves and in particular Riemann surfaces. 
We find that 
our states correspond to Riemann surfaces with genus up to seven,
depending on their symmetry groups, which are all subgroups of $SO(6)$.
In section 5 we provide details concerning the geometrical origin of
the supersymmetric states in five dimensions from a ten-dimensional 
point of view using various distributions of D3-branes in type-IIB
supergravity. This approach yields explicit expressions for the
metric and the scalar fields, and it can be viewed as complementary to 
the algebro-geometric classification of section 4 in terms of 
Riemann surfaces.
In section 6 we consider massless scalar and graviton fluctuations
propagating on our backgrounds. We formulate the problem equivalently
as a Schr\"odinger equation in one dimension and compute 
the potential in some cases of particular interest. 
We also note intriguing connections of these potentials to 
Calogero models and various elliptic generalizations thereof. 
Finally, we end the paper with section 7 where we present our conclusions and
some directions for future work.

\section{Elements of five-dimensional gauged supergravity}

${\cal N}=8$ supergravity in five dimensions involves 42 scalar fields 
parametrizing the non-compact coset space $E_{6(6)}/USp(8)$
that describes their couplings in 
the form of a non-linear $\s$-model \cite{cremmer}.
In five-dimensional 
gauged supergravity the global 
symmetry group $E_{6(6)}$ breaks into an $SO(6)$ subgroup which corresponds
to the gauge symmetry group of the resulting theory, and 
a non-trivial potential develops \cite{PPN,GRW}. 
In the framework of the AdS/CFT correspondence \cite{Maldacena,Witten,GKP}
the supergravity scalars 
represent the couplings of the marginal and relevant chiral primary operators 
of the ${\cal N}=4$ supersymmetric Yang--Mills theory in four dimensions. 
The invariance of the theory with respect to the gauge
group, as well as the $SL(2,\IR)$ 
symmetry inherited from type-IIB supergravity 
in ten dimensions, restricts the scalar potential to depend on $42-15-3=24$
invariants of the above groups. However, it seems still 
practically impossible to deal with such a general potential. 
In this paper we restrict attention to the scalar subsector
corresponding to the symmetric traceless
representation of $SO(6)$, which parametrizes the coset $SL(6,\IR)/SO(6)$, 
and set all other fields (except the metric) equal to zero. 
In this sector we will be able to find explicitly the general solution of the 
classical equations of motion that preserves supersymmetry.

The Lagrangian for this particular coupled gravity-scalar sector includes 
the usual 
Einstein--Hilbert term, the usual kinetic term for the scalars as well 
as their potential 
\be
{\cal L} = {1\ov 4} {\cal R} - \ha \sum_{i=1}^{5} (\del \a_i)^2 - P\ .
\label{lag1}
\ee
A few explanations concerning the scalar-field part of this action are in
order. It has been shown that in this subsector the scalar potential $P$ 
depends on the symmetric matrix $SS^T$ only, where $S$ is an element of 
$SL(6,\IR)$ \cite{GRW} (for a recent discussion see also \cite{DZ,FGPW2}). 
Diagonalization of this matrix yields a form that 
depends only on five scalar fields. 
It is convenient, nevertheless, to represent this sector in terms of six 
scalar fields $\beta_i$, $i=1,2,\dots , 6$ as \cite{FGPW2}
\be
P= -{1\ov 8 R^2} \left( (\sum_{i=1}^6 e^{2 \beta_i})^2 
- 2 \sum_{i=1}^6 e^{4 \beta_i} \right)\ ,
\label{ppotenti}
\ee
where 
\be
\pmatrix{ \b_1 \cr \b_2 \cr \b_3 \cr \b_4 \cr \b_5 \cr \b_6} = 
\pmatrix{ 1/\sqrt{2} & 1/\sqrt{2}& 1/\sqrt{2} & 0 & 1/\sqrt{6}\cr
1/\sqrt{2} & -1/\sqrt{2}& - 1/\sqrt{2} & 0 & 1/\sqrt{6}\cr
- 1/\sqrt{2} & - 1/\sqrt{2}&  1/\sqrt{2} & 0 & 1/\sqrt{6}\cr
- 1/\sqrt{2} & 1/\sqrt{2}& - 1/\sqrt{2} & 0 & 1/\sqrt{6}\cr
0  & 0 &0  & 1 & -\sqrt{2/3}\cr
0  & 0 &0  &- 1 & -\sqrt{2/3}}
\pmatrix{ \a_1 \cr \a_2 \cr \a_3 \cr \a_4 \cr \a_5}\ .
\label{beaa}
\ee
Note that the $6\times 5$ matrix that relates the auxiliary scalars 
$\b_i$ with the $\a_i$'s is not unique; it only has to satisfy the 
condition $\sum_i\b_i =0$. The choice in \eqn{beaa} is particularly
useful for certain computational purposes. It also has the property that 
if the fields $\b_i$ are canonically normalized, the five independent 
scalar fields $\a_i$ will be canonically normalized as well, 
i.e. $\sum_{i=1}^6 (\del \b_i)^2 = 2 \sum_{i=1}^5 (\del \a_i)^2$.

The form of the kinetic term for the scalars in \eqn{lag1} suggests that the 
metric in the corresponding coset space is taken to be $\d^{ij}$. 
This was explicitly shown for the case of only one scalar field in
\cite{DZ} and the general result was quoted without detailed explanation 
in \cite{FGPW2}.
One can generally prove this statement by first realizing that 
the kinetic term of these scalars can depend on two type of terms, 
namely ${\rm Tr}(\del_\m S S\inv) {\rm Tr}(\del^\m S S\inv)$ and 
${\rm Tr}(\del_\m S S\inv \del^\m S S\inv)$. Since $\del_\m S S\inv$
belongs to the algebra of $SL(6,\IR)$ the first term is zero 
because of the traceless condition. The second term gives a result 
proportional to $\sum_{i=1}^6 (\del \b_i)^2 = 2 \sum_{i=1}^5 (\del \a_i)^2$,
thus showing that the scalar kinetic term in 
\eqn{lag1} has indeed the above form.
The equations of motion follow by varying the action \eqn{lag1}
with respect to the five-dimensional metric and the scalar fields.
Using the metric $G_{MN}$, we have 
\ba
{1\ov 4} R_{MN} & =&  \ha \sum_{i=1}^5 \del_M \a_i \del_N \a_i + {1\ov 3} 
G_{MN} P  \ ,
\nonumber \\
D^2 \a_i & =&  {\del P \ov \del \a_i}\ .
\label{eqs32}
\ea
There is a maximally supersymmetric solution of the above equations
that preserves all 32 supercharges,
in which all scalar fields are set zero and the metric is that of 
$AdS_5$ space.
Then, the potential in \eqn{ppotenti} becomes $P=-3/R^2$ and equals 
by definition to the negative cosmological constant of the theory.
This defines the length scale $R$ that will be used in the following. 

The coupled system of non-linear differential equations \eqn{eqs32} is 
in general difficult to solve. In this paper we will be interested 
in solutions preserving four-dimensional Poincar\'e invariance $ISO(1,3)$.
Hence, we make the following ansatz for the five-dimensional metric
\be
ds^2 = e^{2 A}(\eta_{\m\n} dx^\m dx^\n + dz^2)\ ,
\label{aans}
\ee
where $\eta_{\m\n}={\rm diag}(-1,1,1,1)$ is the 
four-dimensional Minkowski metric 
and the conformal factor $e^{2A}$, as well as the scalar fields $\a_i$, depend 
only on the variable $z$.
In addition, we demand that our 
solutions preserve supersymmetry. The corresponding Killing spinor
equations, arising from the supersymmetry transformation rules for the 
8 gravitinos and the 42 spin-$1/2$ fields, give rise to the first 
order equations \cite{FGPW2}
\be
A^\prime  = {2\ov 3 R} e^A  W\ ;\qq 
\a^\prime_i  =  -{1\ov R} e^A {\del W\ov \del \a_i}\ ,\qq i =1,2,\dots, 5\ ,
\label{hjd1}
\ee
where 
\be
W=-{1\ov 4}\sum_{i=1}^6 e^{2 \b_i}\ ,
\label{h2d1}
\ee
and the derivative is taken with respect to the coordinate $z$.
It is straightforward to check that all supersymmetric solutions satisfying 
the first order equations \eqn{hjd1} also satisfy the second order equations
\eqn{eqs32}. In doing so, it is convenient to use the alternative
expression for the potential, instead of \eqn{ppotenti}, 
\be
P = {1\ov 2 R^2} \sum_{i=1}^5 \left(\del W\ov \del \a_i\right)^2 
- {4\ov 3 R^2} W^2\ .
\label{alt1}
\ee

\section{The general supersymmetric solution}

\setcounter{equation}{0}

We begin this section with the construction of the most general solution of 
the non-linear system of equations \eqn{hjd1} 
and discuss some of the general properties of the corresponding 
supersymmetric configurations.
We also show how our solution can be lifted to ten dimensions in the context
of type-IIB supergravity.

\subsection{Five-dimensional solutions}
It might still seem difficult to find solutions of the coupled
system of equations \eqn{hjd1} at first sight, 
due to non-linearity.
It turns out, however, that this is 
not the case, but instead it is possible to find the most general solution. 
In order to proceed further,
we first compute the evolution of the auxiliary scalar fields $\b_i$.
Using \eqn{beaa} and \eqn{hjd1} we find
\be
\b_i^\prime =  {e^A\ov R}\left({2\ov 3} W + e^{2 \b_i} \right)
 =  A^\prime + {1\ov R}  e^{2 \b_i + A}
\ ,\qq i=1,2,\dots , 6\ ,
\label{eh1}
\ee
where for the last equality we have used the first equation in \eqn{hjd1}.
This substitution results into six decoupled first order equations for the
$\b_i$'s which can be easily integrated, as we will soon demonstrate. 
Of course, after deriving 
the explicit solution for the $\b_i$, we also have to check the
self-consistency of this substitution.

Let us reparametrize the function $A(z)$ in terms of an 
auxiliary function $F(z/R^2)$ as follows
\be
e^A = {1\ov R} (-F^\prime/2)^{1/3}\ .
\label{ai1}
\ee 
We have included a minus sign in this definition since, according to the 
boundary conditions that we will later choose, $F$ will be a decreasing 
function of $z$.
Then, according to this ansatz, the general solution of \eqn{eh1} 
is given by 
\be
e^{2 \b_i} = {(-F^\prime/2)^{2/3}\ov F-b_i}\ ,
\qq i=1,2,\dots , 6\ ,
\label{hja1}
\ee
where the prime denotes here the derivative with respect to the 
argument $z/R^2$.
The $b_i$'s are six constants of integration,
which, sometimes is convenient to order as
\be
b_1\geq b_2\geq \dots \geq b_6\ ,
\label{hord1}
\ee
without loss of generality.
Note that we may fix one combination of them to an arbitrary
constant value because \eqn{ai1} determines the 
function $F$ up to an additive constant. 
Also, since the sum of the $\b_i$'s
is zero, we find that 
the function $F$ has to satisfy the differential equation
\be
(F^\prime)^2 =4 \prod_{i=1}^6 (F-b_i)^{1/2}\ ,
\label{jds1}
\ee
which thus contains all the information about the supersymmetric 
configurations and provides a non-trivial algebraic constraint.
Using \eqn{ai1}, \eqn{hja1} and \eqn{jds1} one may easily check that
the first equation in \eqn{hjd1} is also satisfied. 
If we insist on presenting the solution in the conformally flat form 
\eqn{aans} the differential equation \eqn{jds1} needs to be solved 
to obtain $F(z/R^2)$. 
This will be studied in detail in section 4, as it is a necessary step
for investigating  the massless scalar and graviton fluctuations in section 6.

At the moment we
present our general solution in an alternative coordinate system, where 
$F$ is viewed as the independent variable.
Indeed, using \eqn{jds1}, we obtain for the metric 
\be
ds^2  =  {f^{1/6}\ov R^2} \eta_{\m\n}dx^\m dx^\n 
+ {R^2 \ov 4 f^{1/3}} dF^2\ ; \qq 
f = \prod_{i=1}^6 (F-b_i)\ ,
\label{fh3}
\ee
whereas the expression for the scalar fields in \eqn{hja1} becomes
\be
e^{2 \b_i} = {f^{1/6}\ov F-b_i}\ ,
\qq i=1,2,\dots , 6\ .
\label{hja2}
\ee
When the constants $b_i$ are all equal, our solution becomes nothing 
but $AdS_5$ with all scalar fields turned off to zero.
In the opposite case, 
when all constants $b_i$ are unequal from one another, there is
no continuous subgroup of $SO(6)$ preserved by our solution.
If we let some of the $b_i$'s to coincide we restore various 
continuous subgroups of $SO(6)$ accordingly.
As for the five scalar fields $\a_i$, they can be found using \eqn{beaa}
\ba
\a_1 & = & {1\ov 2\sqrt{2}} \left( \b_1 +  \b_2 - \b_3 -\b_4\right)\ ,
\nonumber\\
\a_2 & = & {1\ov 2\sqrt{2}} \left( \b_1 +  \b_4 - \b_2 -\b_3\right)\ ,
\nonumber\\
\a_3 & = & {1\ov 2\sqrt{2}} \left( \b_1 +  \b_3 - \b_2 -\b_4\right)\ ,
\label{anti1}\\
\a_4 & = & \ha (\b_5-\b_6) 
\nonumber\\
\a_5 & = & -\sqrt{3\ov 8} (\b_5+\b_6) \ .
\nonumber
\ea

Note that imposing the reality condition on the scalars in \eqn{hja2}
restricts the values of $F$ to be larger that the maximum of the constants 
$b_i$, which according to the ordering in \eqn{hord1} means that $F\ge b_1$.
For $F\gg b_1$ the scalars tend to zero and $f\simeq F^6$, in which case 
the metric
in \eqn{fh3} approaches $AdS_5$ as expected; put differently, in this limit 
$F\simeq 1/z$ close to $z=0$ that is taken as the origin of the 
$z$-coordinate.
For intermediate values of $F$ we have a flow in the five-dimensional 
space spanned by all scalar fields $\b_i$. 
In general we may have $b_1=b_2=\dots = b_n$, 
with $n\leq 6$, when $b_1$ is $n$-fold degenerate.
In this case, the solution preserves an $SO(n)$ subgroup of $SO(6)$ 
and the flow is actually taking place in $6-n$ dimensions. 
On the other hand, let us consider
the case when $F$ approaches its lower value $b_1$. 
Then, the scalars in \eqn{hja2} are approaching 
\ba
e^{2\b_i}\simeq  \left \{ \begin{array} {ccc}
 f_0^{1/6} (F-b_1)^{(n-6)/6}\ , \quad & {\rm for} \ \ & i=1,2,\dots , n \\ 
\\
{f_0^{1/6}\ov b_1-b_i} (F-b_1)^{n/6}\ ,\quad & {\rm for} \  \ & 
i=n+1,\dots , 6 \\
\end{array} \right \} \ ,
\label{jkef1}
\ea
where $f_0=\prod_{i=n+1}^6 (b_1-b_i)$. Consequently, we have a 
one-dimensional flow in this limit since the scalar fields $\b_i$ can be 
expressed in terms of a single (canonically normalized) scalar $\a$, as 
\ba
{\bf \b}  & \simeq & {1\ov \sqrt{3 n(6-n)}} (n-6,\dots, n-6,n,\dots,n)\ \a\ ,
\nonumber\\
\a & \simeq  & {\sqrt{n(6-n)}\ov 4 \sqrt{3}} \ln(F-b_1)\ .
\label{fjk23}
\ea
It is also useful to find the limiting form of the metric \eqn{fh3} when
$F\to b_1$. Changing the variable to $\r$ as
\be
F= b_1 + \left({(6-n) f_0^{1/6} \r \ov 3 R}\right)^{6\ov 6-n} \ ,
\label{jdr23}
\ee
the metric \eqn{fh3} becomes for $\r\to 0^+$ 
\be
ds^2 \simeq d\r^2 + \left(\left(6-n\ov 3\right)^n {f_0\ov  
R^{12-n}}\right)^{1\ov 6-n}
\r^{n\ov 6-n} \eta_{\m\n} dx^\m dx^\n \ .
\label{jksd1}
\ee
Hence, at $\r=0$ (or equivalently at $F=b_1$) there is a naked singularity
which has an interpetation, as we will see later in the 
ten-dimensional context,
as the location of a distribution of D3-branes.
It is instructive to compare this with the singular behaviour of non-conformal 
non-supersymmetric solutions found in \cite{KS1}. 
A similar naked singularity was found there, but the 
corresponding metric near the singularity had a power law behaviour in
$\r$ with exponent equal to $1/2$, which coincides
with the result in \eqn{jksd1} only for $n=2$.

\subsection{Type-IIB supergravity origin}

It is possible to lift our solution with metric and scalars
given by \eqn{fh3} and \eqn{hja2} to a 
supersymmetric solution of type-IIB supergravity, where only the metric and
the self-dual five-form are turned on. This proves
that our five-dimensional solution is a true compactification of 
type-IIB supergravity on $S^5$. This is not a priori 
obvious because unlike the case of the $S^7$ 
compactification of eleven-dimensional supergravity to four dimensions
\cite{witnik},
there is no general proof that the full non-linear five-dimensional gauged
supergravity action can be fully encoded into the action or equations
of motion of the type-IIB supergravity for the $S^5$ compactification.
However, there is a lot of evidence that this is indeed the case and 
our result gives further support in its favour.

We will show that the ten-dimensional metric corresponds to the 
gravitational field of a large number of D3-branes in the field theory limit
with a special continuous 
distribution of branes in the transverse six-dimensional
space. Namely, the metric has the form
\be
ds^2 = H_0^{-1/2} \eta_{\m\n} dx^\m dx^\n + H_0^{1/2} (dy_1^2+dy_2^2+\dots +
dy_6^2)\ ,
\label{d33}
\ee
where $H_0$ is a harmonic function (yet to be determined)
in the six-dimensional space
transverse to the brane parametrized by the $y_i$ coordinates.
However, instead of being asymptotically flat, the metric \eqn{d33}
will become asymptotically 
$AdS_5\times S^5$ for large radial distances (or equivalently in the UV
region using the terminology of the AdS/CFT correspondence).
The ten-dimensional dilaton field is constant, i.e. $e^\Phi = g_s =
{\rm const.}$ and, as usual, the self-dual five-form is turned on.
Under these conditions, the ten-dimensional solution breaks half of the 
maximum number of supersymmetries (see, for instance, \cite{KaKou}).

We proceed further by first performing the coordinate change in \eqn{d33}
\be
y_i = R e^{A-\b_i} \hat x_i = (F-b_i)^{1/2} \hat x_i\ ,\qq i=1,2,\dots , 6\ ,
\label{ejh1}
\ee
where the $\hat x_i$'s define a unit five-sphere, i.e. they
obey $\sum_{i=1}^6 \hat x_i^2 =1$.
Various 
convenient bases for these unit vectors can be 
chosen, depending on the particular applications that will be presented later.
It can be shown that the flat six-dimensional metric
in the transverse part of the brane metric \eqn{d33} can be written as
\be
 \sum_{i=1}^6 dy_i^2 = R^2 e^{2 A} d\hat \s^2 + 
{e^{-2 A}\ov 4 R^2} \sum_{i=1}^6 e^{2 \b_i} \hat x_i^2\ dF^2\ ,
\label{d34}
\ee
where the line element $d\hat \s^2$ defines the metric of a 
deformed five-sphere given by
\be
d\hat \s^2 = \sum_{i=1}^6 e^{-2 \b_i} (d\hat x_i)^2 \ ,\qq
\det \hat g = {\rm Vol}({S^5}) \sum_{i=1}^6 e^{2 \b_i} \hat x_i^2 \ . 
\label{kjf1}
\ee
For later use, 
we have also written the expression for the determinant of the
deformed five-sphere in \eqn{kjf1}. In computing this determinant 
we have used the fact that the 
sum of the $\b_i$'s is zero. Note that a similar expression also holds for 
a general $n$-sphere.

The harmonic function $H_0$ is determined by 
comparing the massless scalar equation $\Box_{10}\Phi=0$
for the ten-dimensional metric \eqn{d33}
with the equation arising 
using the five-dimensional metric \eqn{aans}, i.e. $\Box_{5}\Phi=0$.
In both cases one makes the ansatz that the solution 
does not depend on the sphere coordinates, i.e. $\Phi= e^{i k \cdot x}
\phi(z)$.
Since the solutions for the scalar $\Phi$ should be the same in 
any consistent trancation of theory, 
the resulting second order ordinary differential
equations should be identical. 
A comparison of terms proportional to $\phi(z)$
determines the function $H_0$ as follows,
\be
H_0^{-1}  = {1 \ov R^4}  f^{1/2} \sum_{i=1}^6 {\hat x_i^2\ov F-b_i}
 = {1 \ov R^4}  f^{1/2} \sum_{i=1}^6 {y_i^2\ov (F-b_i)^2}\ ,
\label{dhj1}
\ee
where in the second equality 
the harmonic function $H_0$ has been expressed in terms 
of the transverse coordinates $y_i$. Comparison of the terms
proportional to the first and second derivative of $\phi(z)$ yields,
using the expression for $\det \hat g$ in \eqn{kjf1}, an identity 
and provides no further information.
The coordinate $F$ is determined in terms of the transverse coordinates
$y_i$ as a solution of the algebraic equation
\be
\sum_{i=1}^6  {y_i^2\ov F-b_i} =1\ .
\label{jk4}
\ee
This is a sixth order algebraic 
equation for general choices of the constants $b_i$, 
and its solution cannot be written in closed form.
However, this becomes possible when some of the $b_i$'s coincide
in such a way that the
degree of \eqn{jk4} is reduced to four or less. Even then, the
resulting expressions are not very illuminating and we will refrain from 
presenting them except in the simplest case in section 5 below.

The corresponding D3-brane solution that is asymptotically flat
is obtained by replacing $H_0$ in \eqn{d33} by $H=1+H_0$. 
Then, in this context, the length parameter 
$R$ has a microscopic interpretation using 
the string scale $\a'$, the string coupling 
$g_s$, and the (large) number of D3-branes $N$, as 
$R^4 = 4 \pi g_s N \a'^2$.

In the rest of this section we
demonstrate for completeness the proof that the function
$H_0$, as defined in \eqn{dhj1}, is indeed harmonic in the six-dimensional 
transverse space spanned by $y_i$, $i=1,2,\dots , 6$. This is not a 
trivial check since $F$ that appears in \eqn{dhj1} is itself a function of
the transverse space coordinates $y_i$ due to the condition \eqn{jk4}.
For notational convenience we define the functions
\be
A_m= \sum_{i=1}^6 {y_i^2\ov (F-b_i)^m}\ ,\qq B_m=\sum_{i=1}^6 {1 \ov 
(F-b_i)^m}\ .
\label{jd23}
\ee
Then, using \eqn{jk4} we determine the derivative of the 
function $F(y)$ 
\be
\del_i F = 2 {y_i \ov A_2 (F-b_i)}\ .
\label{lll3}
\ee
Also, the first derivative of $H_0$ with respect to $y_i$ turns out to be
\be
\del_i H_0  =  - f^{-1/2} {B_1 y_i\ov A_2^2 (F-b_i)}
 - 2 f^{-1/2} {y_i \ov A_2^2 (F-b_i)^2}
+ 4 f^{-1/2} {A_3 y_i\ov A_2^3 (F-b_i)} \ .
\label{hjg4}
\ee
Taking the derivative with respect to $y_i$ once more, summing over the free
indices and after some algebraic manipulations, we obtain the desired result
\ba
\sum_{i=1}^6 \del_i^2 H_0 & = & 2 f^{-1/2} \left({B_2\ov A_2^2} - {B_1 A_3\ov
A_2^3}\right) 
\nonumber \\
&& -\ 2 f^{-1/2} \left({B_2\ov A_2^2} - {B_1 A_3\ov A_2^3}\right) +
16 f^{-1/2} \left({A_4\ov A_2^3} - {A_3^2\ov A_2^4}\right) 
\nonumber\\
&& -\ 16 f^{-1/2} \left({A_4\ov A_2^3} - {A_3^2\ov A_2^4}\right) \ = \ 0\ ,
\label{jfh56}\\
\nonumber 
\ea
where the terms appearing in the three different lines above arise from the 
three distinct terms of \eqn{hjg4} respectively.

\section{Riemann surfaces in gauged supergravity}
\setcounter{equation}{0}

In this section we will present the basic mathematical aspects of 
our general ansatz for the supersymmetric conditions of five-dimensional 
gauged supergravity and find the means to obtain explicit solutions
in several cases by appealing to methods of algebraic geometry.
In fact, we will classify all possible solutions according to 
symmetry groups (subgroups of $SO(6)$) and use the uniformization
of algebraic curves that result in this approach for deriving
the corresponding expressions. To simplify matters the parameter $R$
will be set equal to 1, but it can be easily reinstated by appropriate 
scaling in $z$.
 

\subsection{Schwarz--Christoffel transform}

A useful way to think about the differential equation for the 
unknown function $F(z)$ is in the context of complex analysis. 
Suppose that $z$ and $F$ are extended in the complex domain
and let us consider a closed polygon in the $z$-plane, including
its interior, and map it via a Schwarz--Christoffel transformation
onto the upper half $F$-plane. This provides a one-to-one conformal
transformation and it is assumed that $F(z)$
is analytic in the polygon and is 
continuous in the closed region consisting of the polygon together 
with its interior. Considering the behaviour of $dz$ and
$dF$ as the polygon is transversed in the counter-clockwise direction,
we know that the transformation is described as
\be
{dz \over dF} = A (F-b_1)^{-{\varphi}_1 /\pi}
(F-b_2)^{-{\varphi}_2 /\pi} \cdots (F-b_n)^{-{\varphi}_n /\pi} ~,
\ee
where $A$ is some constant that changes by rescaling $F$.
The vertices of the polygon are mapped to the points 
$b_1, b_2, \cdots ,b_n$ on the real axis of the upper complex 
$F$-plane and the exponents ${\varphi}_i$ that appear in the 
transformation are the exterior (deflection) 
angles of the polygon at the corresponding vertices.
When the polygon is closed their sum is 
${\varphi}_1 + {\varphi}_2 + \dots + {\varphi}_n = 2\pi$. 
Of course, without loss
of generality, we may take one point (say $b_n$) to infinity. Letting
$A = B/(-b_n)^{-{\varphi}_n /\pi}$ we see that as $b_n \rightarrow \infty$
the Schwarz--Christoffel transformation becomes
\be
{dz \over dF} = B (F-b_1)^{-{\varphi}_1 /\pi}(F-b_2)^{-{\varphi}_2 /\pi}
\cdots (F-b_{n-1})^{-{\varphi}_{n-1} /\pi} ~,
\ee
where $B$ is another constant factor. To make contact with our problem
we choose $n=7$ and let the angles ${\varphi}_1 = {\varphi}_2 = \cdots = 
{\varphi}_6 = \pi/4$. Then, we arrive at the differential equation
\be
\left({dz \over dF}\right)^4 = B^4(F-b_1)^{-1}(F-b_2)^{-1} \cdots
(F-b_6)^{-1} ~,
\ee
which is the same as the one implied by our ansatz for the general
solution of gauged supergravity (with $B=1/2$).

The solutions of this equation are difficult to obtain in practice
for generic values of the moduli $b_i$. We will investigate
this problem in connection with the theory of algebraic curves in 
$C^2$ and we will see that in many cases explicit solutions can be
given using the theory of elliptic functions. Before 
proceeding further we note that in our formulation we are looking
for the map from the interior of the polygon onto the 
upper half-plane, $F(z)$, and not for the inverse transformation.


\subsection{Symmetries and algebraic curves}

If we extend the variable $z$ to the complex domain, as before, and 
set 
\be
x= 4F(z) ~, ~~~~~~ y=4F^{\prime}(z) ~ , ~~~~~~ \l_i = 4b_i ~,
\ee
the Schwarz--Christoffel differential 
equation will become an algebraic curve in $C^2$,
\be
y^4 = (x-\l_1)(x-\l_2) \dots (x-\l_6) ~.
\ee
This is a convenient formulation for finding solutions of the
supersymmetry equations, but at the end we have to restrict to
real values of $z$ and demand that the resulting 
supergravity fields ${\beta}_i$
are also real. For generic values of the parameters $b_i$, so that 
they are all unequal and hence there is no symmetry in the solution
of five-dimensional gauged supergravity, the genus of the curve can be easily
determined (like in any other case) 
via the Riemann--Hurwitz relation. Recall that for any 
curve of the form
\be
y^m = (x-\l_1)^{\a_1}(x-\l_2)^{\a_2} \dots (x-\l_n)^{\a_n}\ ,
\ee
which is reduced, i.e. the integers $m$ and $\a_i$ have no common factors, 
and all $\l_i$'s are unequal, 
the genus $g$ can be found by first writing the ratios
\be
{\a_1 \over m} = {d_1 \over c_1} ~, ~~ \cdots ~~ , {\a_n \over m} =
{d_n \over c_n} ~; ~~~~~~ {\a_1 + \cdots + \a_n \over m} = {d_0 \over c_0} 
\ee
in terms of relatively prime numbers and then using the relation
\be
g = 1 - m +{m \over 2} \sum_{i=0}^n \left(1-{1 \over c_i}\right) .
\ee
According to this the genus of our surface turns out to be $g=7$
when all $b_i$ are unequal, 
and so it is difficult to determine explicitly the
solution in the general case. However, by imposing some isometries
in the solution of gauged supergravity the genus becomes smaller
and hence the problem becomes more tractable. The presence of isometries
manifests by allowing for multiple branch points in the general form of the
algebraic curve, which in turn degenerates along certain cycles
that effectively reduce its genus. 

Note for completeness that if we had not taken $b_7$ to infinity 
in our discussion of the Schwarz--Christoffel
transformation, we would have had an additional factor $(x-\l_7)^2$ in the
equation of the algebraic curve because ${\varphi}_7 = \pi/2$ instead of
$\pi/4$ that was chosen for the remaining ${\varphi}_i$'s. It can be easily
verified that this does not affect the genus of the curve, as expected
on general grounds.

Next, we enumerate all possible cases with a certain degree of 
symmetry that correspond to various subgroups of $SO(6)$; this 
amounts to various deformations of the round five-sphere, $S^5$, which
is used for the compactification of the theory from 10 to 5 dimensions.
Consequently, this will in principle determine the solution for the
scalar fields in the remaining 5 dimensions as we will 
see later in detail. If all the branch points are different, the $SO(6)$
isometry of $S^5$ will be completely broken, whereas if all of them
coalesce to the same point the maximal isometry $SO(6)$ will be 
manifestly present. The classification is presented below in an order of
increasing symmetry or else in decreasing values of $g$.

\noindent
(1) \underline{SO(2)}: It corresponds to setting two of the $\l_i$
equal to each other and the remaining are all unequal. The curve becomes
\be
y^4 = (x-\l_1)(x-\l_2)(x-\l_3)(x-\l_4)(x-\l_5)^2
\ee
and its genus turns out to be $g=5$.

\noindent
(2) \underline{SO(3)}: It corresponds to setting three of the 
$\l_i$ equal and all other remain unequal. The curve becomes
\be
y^4 = (x-\l_1)(x-\l_2)(x-\l_3)(x-\l_4)^3
\ee
and the genus turns out to be $g=4$.

\noindent
(3) \underline{$SO(2) \times SO(2)$}: In this case two pairs of $\l_i$
are mutually equal and the remaining two parameters are unequal. The
curve becomes
\be
y^4 = (x-\l_1)(x-\l_2)(x-\l_3)^2(x-\l_4)^2
\ee
and its genus is $g=3$.

\noindent
(4) \underline{$SO(3) \times SO(2)$}: In this case three $\l_i$ are
equal and another two are also equal to each other. The curve becomes
\be
y^4 = (x-\l_1)(x-\l_2)^2(x-\l_3)^3
\ee
and its genus is $g=2$. Therefore we know that it can be cast into
a manifest hyper-elliptic form by introducing appropriate bi-rational 
transformations of the complex variables.

\noindent
(5)  \underline{SO(4)}: It corresponds to setting four $\l_i$ equal to
each other and the other two remain unequal. The curve becomes
\be
y^4 = (x-\l_1)(x-\l_2)(x-\l_3)^4
\ee
and its genus is $g=1$. It can also be cast into a manifest (hyper)-elliptic
form as we will see shortly.

\noindent
(6) \underline{$SO(2) \times SO(2) \times SO(2)$}: It corresponds to 
three different pairs of mutually equal $\l_i$, but in this case
the curve is not irreducible, since $y^4 = (x-\l_1)^2(x-\l_2)^2(x-\l_3)^2$.
The reduced form is 
\be
y^2 = (x-\l_1)(x-\l_2)(x-\l_3)
\ee
and clearly has genus $g=1$ as it is written directly in (hyper)-elliptic
form.

\noindent
(7) \underline{$SO(3) \times SO(3)$}: In this case we have two groups of 
triplets with equal values of $\l_i$. The curve becomes
\be
y^4 = (x-\l_1)^3(x-\l_2)^3
\ee
and its genus is $g=1$. It can also be cast into a manifest (hyper)-elliptic
form.

\noindent
(8) \underline{SO(5)}: In this case five $\l_i$ are equal to each other
and the last remains different. The curve becomes
\be
y^4 = (x-\l_1)(x-\l_2)^5
\ee
and its genus is also $g=1$ as before.

\noindent
(9) \underline{$SO(4) \times SO(2)$}: It corresponds to separating the 
$\l_i$ into four equal and another two equal parameters. The curve becomes
$y^4 = (x-\l_1)^2(x-\l_2)^4$, but it is not irreducible. The reduced 
form is 
\be
y^2 = (x-\l_1)(x-\l_2)^2
\ee
and has genus $g=0$, as it can also be obtained by degenerating a genus 1
surface along its cycles. Therefore, we expect the solution to be given 
in terms of elementary functions.

\noindent
(10) \underline{SO(6)}: This is the case of maximal symmetry in which all
$\l_i$ are set equal to each other. The curve becomes
$y^4 = (x-\l_1)^6$, whose reduced form is
\be
y^2 = (x-\l_1)^3
\ee
and has genus $g=0$ as before.

Of course, when certain cycles contract by letting various branch points
to coalesce, the higher genus surfaces reduce to lower genus and a bigger 
symmetry group emerges in the solutions corresponding to
gauged supergravity. For 
genus $g \leq 2$ one can always transform to a manifest hyper-elliptic form
so that two sheets (instead of four) are needed for picturing the Riemann
surface by gluing sheets together along their branch cuts.  
We will investigate in detail the cases corresponding to genus 0 and
1 surfaces since the solutions can be given explicitly in terms of 
elementary and elliptic functions respectively. Some results about the 
genus 2 case will also be presented. The other cases are more difficult to 
handle in detail even though the general form of the solution is known  
implicitly for all $g$ according to our ansatz.


\subsection{Genus 0 surfaces}

There are two genus 0 surfaces according to the previous discussion, namely
the curve $y^2 = (x-\l_1)^3$ for the isometry group $SO(6)$ and the curve
$y^2 = (x-\l_1)(x-\l_2)^2$ for the isometry group $SO(4) \times SO(2)$. 
According to algebraic geometry every irreducible curve $f(x,y) = 0$
with genus 0 is representable as a unicursal curve (straight line)
\be
w = v
\ee
by means of a bi-rational transformation $x(v,w)$, $y(v,w)$ and 
conversely $v(x,y)$, $w(x,y)$. In our two examples the underlying
transformations are summarized as follows:

\noindent
(a) \underline{SO(6)}: We have
\be
x = vw + \l_1 ~, ~~~~~~ y = vw^2 
\ee
and conversely
\be
v = {(x-\l_1)^2 \over y} ~, ~~~~~~ w = {y \over x-\l_1} ~.
\ee

\noindent
(b) \underline{$SO(4) \times SO(2)$}: We have
\be
x = vw + \l_1 ~, ~~~~~~ y = w(vw + \l_1 - \l_2) 
\ee
and conversely
\be
v = {(x-\l_1)(x-\l_2) \over y} ~, ~~~~~~ w = {y \over x-\l_2} ~.
\ee
Of course, the first curve arises as special case of the second for
$\l_2 \rightarrow \l_1$.

For general $\l_1$ and $\l_2$ we may use $u$ as a (trivial) uniformizing
complex parameter for the unicursal curve, i.e. $v=u=w$. Then, the 
expressions for $x$ and $y$ yield
\be
4F(z) = u^2 + 4b_1 ~, ~~~~~~ 4{dF(z) \over dz} = u\left(u^2 + 4(b_1 - b_2)
\right) ,
\ee
where we have taken into account the rescaling $x=4F(z)$, $y=4F^{\prime}(z)$,
$\l_i = 4b_i$ that was introduced earlier. So we can determine $u$ as a 
function of $z$ by simple integration since
\be
{du \over dz} = {1 \over 2}\left(u^2 + 4(b_1-b_2)\right) .
\ee
In fact there are three different 
cases for generic values of $b_1$ and $b_2$. Choosing appropriately
the integration constant, so that the resulting conformal factor
$e^{2A(z)}$ will behave like $1/z^2$ as $z \rightarrow 0$, we have:
\ba
({\rm i}) ~~~~~u & = & - 2\sqrt{b_1 - b_2} {\rm cot}\left(\sqrt{b_1-b_2}
z\right), ~~~~ {\rm for} ~~ b_1>b_2 ~,\\
({\rm ii}) ~~~~ u & = & -2\sqrt{b_2 - b_1} {\rm coth}\left(\sqrt{b_2-b_1}
z\right), ~~~~ {\rm for} ~~ b_2>b_1 ~,\\
({\rm iii}) ~~~ u & = & -{2 \over z} ~, ~~~~ {\rm for} ~~ b_1=b_2 ~.
\ea
The first two cases correspond to the $SO(4) \times SO(2)$ isometry
and they are
obtained by analytic continuation from one other, 
depending on the size of the $b_i$'s,
whereas the last case has $SO(6)$ isometry. 
Here, we do not assume any given ordering among the $b_i$'s.
As for the functions $F(z)$ we have respectively
\be
F(z) = (b_1-b_2) {\rm cot}^2 \left(\sqrt{b_1-b_2}z\right) + b_1 ~, ~~~~
(b_2-b_1) {\rm coth}^2 \left(\sqrt{b_2-b_1}z\right) + b_1 ~, ~~~~
{1 \over z^2} + b_1 ~.
\ee

Then, the expression for the conformal factor of the metric is
\ba
({\rm i}) ~~~~~ e^{2A(z)} &=& (b_1 -b_2) {{\rm cos}^{2/3}\left(
\sqrt{b_1 -b_2} z\right) \over {\rm sin}^2\left(\sqrt{b_1 -b_2} z\right)} ~,
~~~~ {\rm for} ~~ b_1>b_2 ~,
\label{a42}\\ 
({\rm ii}) ~~~~ e^{2A(z)}&=& (b_2 -b_1) {{\rm cosh}^{2/3}\left(
\sqrt{b_2 -b_1} z\right) \over {\rm sinh}^2 \left(\sqrt{b_2 -b_1} z\right)} ~,
~~~~ {\rm for} ~~ b_2>b_1 ~,
\label{a42a}\\
({\rm iii}) ~~~ e^{2A(z)}&= &{1 \over z^2} ~, ~~~~ {\rm for} ~~ b_1 =b_2 ~,
\ea
which indeed behaves as $1/z^2$ in all three cases for $z \rightarrow 0$.

The solution for the scalar fields 
${\beta}_i(z)$ of five-dimensional gauged supergravity
follows by simple substitution into our ansatz. We have explicitly
in each case
\ba
&({\rm i}) &~~~~  e^{2\beta_1(z)}  =  e^{2\beta_2(z)} = 
{1 \over {\rm cos}^{4/3}\left(\sqrt{b_1 -b_2}z\right)} ~, \nonumber\\
&& \phantom{xx} e^{2\beta_3(z)} = e^{2\beta_4(z)} = e^{2\beta_5(z)} = e^{2\beta_6(z)} =
{\rm cos}^{2/3}\left(\sqrt{b_1-b_2}z\right) ,
\label{b42}\\
&({\rm ii})& ~~~~  e^{2\beta_1(z)} =  e^{2\beta_2(z)} = 
{1 \over {\rm cosh}^{4/3}\left(\sqrt{b_2 -b_1}z\right)} ~,\nonumber\\
&& \phantom{xx}
 e^{2\beta_3(z)} =  e^{2\beta_4(z)} = e^{2\beta_5(z)} = e^{2\beta_6(z)} =
{\rm cosh}^{2/3}\left(\sqrt{b_2-b_1}z\right) ,
\label{b42b}\\
&({\rm iii})& ~~~  e^{2\beta_i(z)}  = 1 ~, ~~~~~~ i=1,\dots , 6 \ .
\ea 

Equally well we could have transformed the genus 0 curves 
into the quadratic form
$Y^2 = 1 - X^2$ using the following transformation for the curve
$y^2 = (x-\l_1)(x-\l_2)^2$
\be
x = {1 + X \over 1-X} + \l_1 ~, ~~~~~~ y = {Y \over 1-X} 
\left({1+X \over 1-X} + \l_1 - \l_2 \right)
\ee
and conversely
\be
X = {x - \l_1 - 1 \over x - \l_1 + 1} ~, ~~~~~~ Y = 
{2y \over (x-\l_1 +1)(x-\l_2)} ~.
\ee
In this case we can use another uniformizing complex parameter $u$, so that
$X={\rm sin}u$ and $Y={\rm cos}u$, and proceed as above.  
Either way, the uniformization problem is solved in terms of elementary
functions, which in turn determine the function $F(z)$ every time and hence
the particular supersymmetric solutions of five-dimensional
gauged supergravity.


\subsection{Genus 1 surfaces}

Recall first that given a genus 1 algebraic curve in its Weierstrass form
\be
w^2 = 4v^3 - g_2v - g_3
\ee
the uniformization problem is solved by introducing the Weierstrass function
${\cal P}(u)$ and its derivative ${\cal P}^{\prime}(u)$ with respect to
a complex parameter $u$. Then, $v={\cal P}(u)$ and
$w={\cal P}^{\prime}(u)$ in which case the Weierstrass function 
satisfies the time independent KdV equation 
${\cal P}^{\prime \prime \prime}(u) -12 {\cal P}(u){\cal P}^{\prime}(u) =0$.
The two periods of the elliptic curve are denoted by $2{\omega}_1$
and $2{\omega}_2$ and the Weierstrass function is double periodic
with respect to them. Also the values of the Weierstrass function at the
half-periods coincide with the roots of the algebraic equation
$4v^3 -g_2v -g_3 = 0$, namely $e_1 = {\cal P}({\omega}_1)$, 
$e_2 = {\cal P}({\omega}_1 + {\omega}_2)$ and $e_3 = {\cal P}({\omega}_2)$.
Conversely, given the differential equation
\be
\left( {dG(z) \over dz}\right)^2 = 4G^3(z) - g_2 G(z) - g_3
\ee
the general solution is described in terms of the Weierstrass function
$G(z) = {\cal P}(\pm z + a)$, where $a$ is the constant of integration
and $g_2$, $g_3$ are related as usual to the periods of the elliptic
curve. This may be seen by taking a new dependent variable $u$ defined
by the equation $G = {\cal P}(u)$, when the differential equation reduces
to $(du/dz)^2 =1$; for this recall that the number of roots of the 
equation ${\cal P}(u) = c$ that lie in any cell depend only on 
${\cal P}(u)$ and not on $c$, which can assume arbitrary values, 
like for any other elliptic function.

We have four different curves with genus 1 that follow from the 
classification that we described above. It is known from algebraic 
geometry that any genus 1 surface is hyper-elliptic (in particular 
elliptic since $g=1$), but we see that only the one
that corresponds to the case of $SO(2) \times SO(2) \times SO(2)$ 
isometry is essentially written in such form with roots $e_1 = \l_1$,
$e_2 = \l_2$ and $e_3 = \l_3$ (when $\l_1 + \l_2 + \l_3 = 0$). 
The other three curves can be 
transformed to $w^2 = 4v^3 -g_2v -g_3$ for appropriately chosen 
coefficients $g_2$ and $g_3$ provided that one performs the necessary
bi-rational transformations of the complex variables
$v(x, y)$, $w(x, y)$ and conversely $x(v,w)$, $y(v, w)$. 
Only then the solution
can be easily deduced from the 
resulting genus 1 curve in its Weierstrass form 
using elliptic functions. This is precisely what we are about to describe in 
the sequel.

Note first that all three curves that correspond to the symmetry groups
$SO(4)$, $SO(3) \times SO(3)$ and $SO(5)$ can be transformed into the same
curve
\be
Y^4 = (X - \l_1) (X - \l_2)
\ee
according to the following bi-rational transformations
\ba
&& X =  x ~, ~~~ Y = {y \over x - \l_3} ~~~~~~ {\rm for} ~~ 
y^4 = (x-\l_1)(x-\l_2)(x-\l_3)^4 ~,\\
&& X  =  x ~ , ~~~ Y = {(x- \l_1)(x-\l_2) \over y} ~~~~~~ {\rm for}
~~ y^4 = (x-\l_1)^3(x-\l_2)^3 ~, \\
&& X  =  x ~ , ~~~ Y = {y \over x-\l_2} ~~~~~~ {\rm for} ~~
y^4 = (x-\l_1)(x-\l_2)^5 \ ,
\ea
respectively. Then, defining 
\be
X- \l_1 = {{\eta}^2 \over \zeta} ~, ~~~~~ Y = {\eta \over \zeta}
\ee
we arrive at the curve
${\eta}^2 (1- {\zeta}^2) = (\l_1 - \l_2){\zeta}^3$ in all three cases.
This simplifies further by defining new variables $V$, $W$ so that
\be
\eta = {W \over V} ~, ~~~~~ {\zeta} + 1 = {1 \over V} ~,
\ee
in which case the curve becomes $W^2 (2V-1) = (\l_1 - \l_2) V(1-V)^3$.
Finally, letting
\be
V = {2v \over \l_2 - \l_1} + {1 \over 2} ~, ~~~~~
W = {1 \over \l_1 - \l_2}{w \over v} \left(v + {1 \over 4}(\l_1 -
\l_2)\right)\ ,
\ee
we obtain the genus 1 curve in its standard Weierstrass form 
\be
w^2 = 4v^3 - g_2v -g_3 ~~~~~~ {\rm with} ~~ g_2 = {1\over 4} 
(\l_1 - \l_2)^2 ~ , ~~ g_3 = 0 
\ee
for all three cases of interest. This is a non-degenerate Riemann surface with 
$g=1$, but it is more special than the $SO(2) \times SO(2) \times SO(2)$
surface since the latter depends on three parameters $\l_1$, $\l_2$ and $\l_3$
instead of the two $\l_1$ and $\l_2$ that appear in the Weierstrass 
form for higher 
(non-abelian) symmetry. Actually, in the present case we have 
${\omega}_2/{\omega}_1 = i$, and so by introducing the modulus of elliptic
integrals, $k$, and its complementary value $k^{\prime}$, one finds
$k = k^{\prime} = 1/\sqrt{2}$. 

We summarize the bi-rational transformations that are needed to transform
each one of the genus 1 surfaces into their Weierstrass forms according
to the symmetry groups of the solutions that they represent:

\noindent
(a) \underline{$SO(2) \times SO(2) \times SO(2)$}: The curve 
$y^2 = (x-\l_1)(x-\l_2)(x-\l_3)$ can be brought into the standard
Weierstrass form $w^2 = 4v^3 - g_2 v - g_3$ with
\ba
g_2& = & {1 \over 36}\left((\l_1 + \l_2 -2\l_3)^2 - 
(\l_2 + \l_3 -2 \l_1)(\l_1 + \l_3 -2 \l_2)\right) ,\\ 
g_3 & = & -{1 \over 432}(\l_1 + \l_2 -2\l_3) 
(\l_2 + \l_3 -2 \l_1)(\l_1 + \l_3 -2 \l_2) \ ,
\ea
using the simple transformation 
\be
y=4w ~, ~~~~~~ x=4v +{1 \over 3}(\l_1 + \l_2 + \l_3) ~.
\ee

\noindent
(b) \underline{$SO(3) \times SO(3)$}: The curve $y^4 = (x-\l_1)^3(x-\l_2)^3$
also transforms into the Weierstrass form $w^2 = 4v^3 - g_2v$ with  
$g_2 = (\l_1 - \l_2)^2 /4$ using
\be
x = \l_1 - {1 \over v} \left(v+{1 \over 4}(\l_1 -\l_2)\right)^2 , ~~~~~
y = {w^3 \over 8v^3}\ ,
\ee
and conversely
\ba
v & = & {1 \over 4}(\l_1-\l_2){y^2 - (x-\l_1)(x-\l_2)^2 \over 
y^2 +(x-\l_1)(x-\l_2)^2} ~,\\
w & = & {1 \over 2}(\l_1 -\l_2){(x-\l_1)(x-\l_2) \over y} 
{y^2 -(x-\l_1)(x-\l_2)^2 \over y^2 + (x-\l_1)(x-\l_2)^2} ~.
\ea

\noindent
(c) \underline{SO(4)}: The curve $y^4 = (x-\l_1)(x-\l_2)(x-\l_3)^4$ transforms
into the Weierstrass form $w^2 = 4v^3 -g_2v$ with $g_2 = (\l_1 -\l_2)^2 /4$
using
\ba
x & = & \l_1 - {1 \over v} \left(v +{1 \over 4} (\l_1 -\l_2)\right)^2 ~,\\
y & = & {w \over 2v} \left(\l_1 - \l_3 - {1 \over v} 
\left(v + {1 \over 4}(\l_1 - \l_2) \right)^2 \right)\ ,
\ea
and conversely
\ba
v & = & {1 \over 4} (\l_2 - \l_1) {y^2 -(x-\l_1)(x-\l_3)^2  \over
y^2 +(x-\l_1)(x-\l_3)^2} ~,\\
w & = & {1 \over 2} (\l_2 -\l_1) {y \over x-\l_3} {y^2 -(x-\l_1)
(x-\l_3)^2 \over y^2 + (x-\l_1)(x-\l_3)^2} ~ .
\ea

\noindent
(d) \underline{SO(5)}: This case arises from $SO(4)$ when 
$\l_3 \rightarrow \l_2$, and so
\be
x = \l_1 -{1 \over v}\left(v+{1 \over 4}(\l_1 -\l_2)\right)^2 , ~~~~~
y = -{w \over 2v^2}\left(v-{1 \over 4}(\l_1 -\l_2)\right)^2 .
\ee 

Note that in the three last cases (b)--(d) one may choose $v = {\cal P}(u)$
and $w = {\cal P}^{\prime}(u)$, where $u$ is the uniformizing parameter
of the same Riemann surface. Thus, the $x$'s ($=4F(z)$) are the same functions
of $u$ in these three cases, given in terms of Weierstrass functions and 
their derivatives, but the $y$'s ($=4F^{\prime}(z)$) are all different 
as can be readily seen. This simply means that the variable $z$ is not
equal to the uniformizing parameter $u$ of the genus 1 curve in its  
Weierstrass form, but rather a more complicated function $u(z)$ that has to be 
found in each case separately by integration (in analogy with what we did 
in the $g=0$ cases). This complication does not
arise for the case (a), since there we can take $z = u$ (more generally 
$(du/dz)^2 = 1$, as we have already seen). For (a) the 
solution has already been very simply expressed in terms of the Weierstrass
functions $x = 4{\cal P}(u) + (\l_1 + \l_2 + \l_3)/3$ 
and $y = 4{\cal P}^{\prime}(u)$, though of 
another Riemann surface with different coefficients $g_2$ and $g_3$.

Next, we take into account the field redefinitions $x = 4F(z)$, 
$y = 4F^{\prime}(z)$, $\l_i = 4b_i$ and solve for $z(u)$ and its inverse
$u(z)$, when this is possible in closed form, 
thus determining $F(z)$ in each case of interest. Of course, the 
elliptic functions that appear, refer to the corresponding curves with
$g_2$ and $g_3$ determined as above. Summarizing the results, including 
some technical details, we have:

\noindent
$\bullet$
\underline{$SO(2) \times SO(2) \times SO(2)$}: We have already seen that
the uniformizing parameter $u$ equals to $z$ and hence
\be
F(z) = {\cal P}(z) + {1 \over 3}(b_1 + b_2 + b_3) 
\label{f222}\ .
\ee
According to this we find
\be
e^{2A(z)} = \left({1 \over 2}{\cal P}^{\prime}(z)\right)^{2/3} \ ,
\label{a222}
\ee
and so the conformal factor of the metric behaves as $1/z^2$ when 
$z$ approaches 0. The solution for the scalar fields of gauged supergravity
follows by substitution into our general ansatz. We have in fact
\ba
&& e^{2\beta_1(z)} = e^{2\beta_2(z)} = {\left({\cal P}^{\prime}(z)/2
\right)^{2/3} \over {\cal P}(z) - e_1} ~, \nonumber\\
&& e^{2\beta_3(z)} = e^{2\beta_4(z)} = {\left({\cal P}^{\prime}(z)/2
\right)^{2/3} \over {\cal P}(z) - e_2} ~, 
\label{s222}\\
&& e^{2\beta_5(z)} = e^{2\beta_6(z)} = {\left({\cal P}^{\prime}(z)/2
\right)^{2/3} \over {\cal P}(z) - e_3} ~, 
\nonumber
\ea
where
\be
e_i = b_i - {1 \over 3}(b_1 + b_2 + b_3) ~, \qq i=1,2,3.
\label{e222}
\ee

\noindent
$\bullet$
\underline{$SO(3) \times SO(3)$}: This is the next simple case
to consider. The 
relation between the differentials $dz$ and $du$ can be found by first
computing $4dF/du$ as a function of $u$; it turns out to be 
$-{{\cal P}^{\prime}}^3(u)/4{\cal P}^3(u)$. Then, using the expression for
$y = 4dF(z)/dz$ we arrive at the simple relation
\be
{du \over dz} = -{1 \over 2}
\ee
and so $u= -z/2$, up to an integration constant that is taken zero. 
Consequently,  
\be
F(z) = b_1 - {1 \over 4{\cal P}(z/2)}\left({\cal P}(z/2) + b_1 - b_2\right)^2,
\ee
which in turn implies the following result for the conformal factor of 
the metric
\be
e^{2A(z)} = \left({{\cal P}^{\prime}(z/2) \over 4 {\cal P}(z/2)}\right)^2
= {1 \ov 4} \left({\cal P}(z/2) - {(b_1-b_2)^2\ov {\cal P}(z/2)}\right)\ .
\label{a33}
\ee
The derivative of the Weierstrass function is taken with respect to 
its argument $z/2$. The conformal factor clearly approaches $1/z^2$ 
as $z \rightarrow 0$, which justifies our choice of the integration 
constant above.

As for the solution corresponding to the scalar fields of gauged
supergravity, we obtain by substitution into our general ansatz the
result
\ba
&& e^{2\beta_1(z)} =e^{2\beta_2(z)} =e^{2\beta_3(z)} = 
-{{\cal P}(z/2) -b_1+b_2 \over {\cal P}(z/2) +b_1-b_2} ~,\nonumber\\
&& e^{2\beta_4(z)} =e^{2\beta_5(z)} =e^{2\beta_6(z)} = 
-{{\cal P}(z/2) +b_1-b_2 \over {\cal P}(z/2) -b_1+b_2} 
\label{b33}\ ,
\ea
which completes the task. 
At this point we add a clarifying remark, which takes into account
the discrete symmetry $x \rightarrow -x$ and $b_i \rightarrow -b_i$
of the underlying algebraic curves. The uniformization that gave rise 
to eq. (4.64) implies that as $z$ ranges from 0 to 
$2 \omega_1$, $F(z)$ ranges from $- \infty$ to $b_2$ (provided that 
$b_1 > b_2$ so that ${\cal P}(\omega_1) \equiv e_1 = b_1 - b_2$).
If one applies the discrete symmetry mentioned above, eq. (4.64)
will change accordingly so that $F(z)$ will range from $+ \infty$  
to $b_1$ (taken as the maximum of the two moduli parameters). This
particular symmetry implies in turn that the expressions for the
scalar fields ${\rm exp}(2 \beta_i)$ get modified by simply changing 
the overall sign according to the defining relation (3.3). 
Hence, despite appearances, the fields $\beta_i(z)$
are real provided that $z$ is real with values 
in the range where $F(z)$
is bigger than the maximum of $b_1$ and $b_2$, as it is usually taken. 

\noindent
$\bullet$
\underline{$SO(5)$}: 
This case is computationally more difficult to
handle. Since $4dF/du = -{{\cal P}^{\prime}}^3 / 4{\cal P}^3(u)$ 
again as a function of $u$, 
we find the following relation between the differentials 
$dz$ and $du$,
\be
{du \over dz} = {1 \over 2}{{\cal P}(u) -b_1 +b_2 \over 
{\cal P}(u) +b_1 -b_2} ~.
\ee
Then, integrating over $u$ we arrive at the formula
\be
{b_2-b_1 \over 2}z = \zeta(u) + {1 \over 2} {{\cal P}^{\prime}(u)
\over {\cal P}(u) -b_1 +b_2} ~ ,
\ee
up to an integration constant that should be determined by the 
asymptotic behaviour $e^{2A(z)} \rightarrow 1/z^2$ as $z$ approaches 0. 
Here $\zeta(u)$ is the Weierstrass zeta-function.
Note that the above expression will somewhat simplify if one uses the
identity
\ba
\zeta(u +\omega_1) - \zeta(\omega_1) &= &
\zeta(u) + {1 \over 2} {{\cal P}^{\prime}(u) \over 
{\cal P}(u) -b_1 +b_2} ~, ~~~~ {\rm for} ~~ b_1>b_2 ~,\\
\zeta(u +\omega_2) - \zeta(\omega_2) &= &
\zeta(u) + {1 \over 2} {{\cal P}^{\prime}(u) \over 
{\cal P}(u) -b_1 +b_2} ~, ~~~~ {\rm for} ~~ b_2>b_1 ~,
\ea
where $\omega_1$ and $\omega_2$ are the half-periods of the curve. 
In either case,
it is not possible to invert the relation and explicitly find $u(z)$ 
in closed form. 

We give the result for the conformal factor
of the metric as a function of $u$,
\be
e^{2A} = {1 \over 4{\cal P}(u)} 
({\cal P}(u) +b_1-b_2)^{1/3}
({\cal P}(u) -b_1 +b_2)^{5/3} ~.
\ee
Similar expressions are obtained for the scalar fields of gauged 
supergravity,
\ba
e^{2\beta_1} &=& - \left({{\cal P}(u) -b_1+b_2 \over 
{\cal P}(u) +b_1-b_2}\right)^{5/3} ~,\nonumber\\
e^{2\beta_2} &= &\cdots = e^{2\beta_6} = - \left(
{{\cal P}(u) + b_1-b_2\over {\cal P}(u)-b_1+b_2}\right)^{1/3} ~.
\ea
Similar remarks apply here for the overall sign appearing in eq. (4.72),
as for the scalar fields of the model $SO(3) \times SO(3)$, using the
discrete symmetry $x \rightarrow -x$, $b_i \rightarrow -b_i$ of the
underlying algebraic curve.

We mention for completeness that as $b_1 \rightarrow b_2$ the Riemann surface
degenerates and one recovers the $SO(6)$ model that was already discussed.
It might seem that this contradicts the relation between $u$ and $z$ at 
first sight, since the left hand side becomes zero 
irrespective of $z$. However, for elliptic 
functions in the degeneration limit $g_2 = g_3 = 0$ we have  
${\cal P}(u) = 1/u^2$ and $\zeta (u) = 1/u$ for all $u$, 
and so the right hand side also becomes zero
irrespective of $u$; hence there is no problem in taking this limit.

\noindent
$\bullet$
\underline{$SO(4)$}: In this situation the calculation becomes even 
more involved. We find that
\be
{du \over dz} = {1 \over 2} {({\cal P}(u) -b_1 + b_2)^2
-4(b_2-b_3){\cal P}(u) \over  
({\cal P}(u) +b_1-b_2)({\cal P}(u)-b_1+b_2)}~,
\ee
but as in the $SO(5)$ case it is still not possible to find explicitly
$u(z)$ in closed form. Besides, the integrals are more difficult to 
perform when $b_2 \neq b_3$ and so the resulting expressions are not 
very illuminating in terms of algebraic geometry. 
We postpone the presentation of the corresponding 
configuration for the next section, where a more geometrical approach
is employed for it.
  

\subsection{Genus 2 surface}

Here we have only one such curve corresponding to the isometry 
group $SO(3) \times SO(2)$, which is described by the algebraic
equation $y^4 = (x-\l_1)(x-\l_2)^2(x-\l_3)^3$. According to 
algebraic geometry it can be brought into a manifest 
hyper-elliptic form by performing appropriate bi-rational
transformations. To achieve this explicitly we consider the
following sequence of transformations: First, let
\be
x= X ~, ~~~~~~ y = {(X-\l_3)(X-\l_2) \over Y} \ ,
\ee
that brings the curve into the form
\be
(X-\l_1)Y^4 = (X-\l_2)^2(X-\l_3) ~.
\ee
The second step consists in performing the transformation
\be
X-\l_2 = {{\eta}^2 \over \zeta} ~, ~~~~~~ Y = {\eta \over \zeta}\ ,
\ee
that transforms it further into the form
\be
{\eta}^2(1-{\zeta}^2) = (\l_1 -\l_2)\zeta + (\l_2 -\l_3){\zeta}^3 ~.
\ee
Next, we introduce $V$ and $W$ so that
\be
\eta = {W \over V} ~, ~~~~~~ \zeta + 1 = {1 \over V}\ ,
\ee
and the algebraic curve simplifies to
\be
W^2(2V-1) = (\l_1-\l_2)V^3(1-V) + (\l_2-\l_3)V(1-V)^3 ~.
\ee
Finally, as last step let us consider
\be
V = v ~, ~~~~~~ W={w \over 2v-1} ~,
\ee
which turns the curve into the desired hyper-elliptic form of genus two
\be
w^2 = v(v-1)(2v-1)\left((\l_3-\l_1)v^2 -2(\l_3-\l_2)v + \l_3 -\l_2
\right)\ ,
\ee
with five distinct roots when all $\l_i$ are different from each 
other. 

Summarizing the sequence of the above operations, which are similar to
the genus 1 examples, we have for the
transformation $x(v, w)$, $y(v,w)$ the final result
\be
x = \l_2 - {(\l_3-\l_1)v^2 -2(\l_3-\l_2)v+\l_3-\l_2 \over 2v-1} ~, ~~~~~ 
y = (\l_1-\l_3){vw \over (2v-1)^2} ~ , 
\ee
whereas for its inverse $v(x,y)$, $w(x,y)$ we have
\be
v = {(x-\l_2)(x-\l_3)^2 \over (x-\l_2)(x-\l_3)^2 + y} ~, ~~~~~
w= {y \over x-\l_3}{(x-\l_2)(x-\l_3)^2 -y^2 \over
(x-\l_2)(x-\l_3)^2 + y^2} \ , 
\ee
and so it is bi-rational, as required. These formulae are useful
for addressing the uniformization problem of the original form
of the curve in terms of theta functions. However, the resulting
solution for gauged supergravity is rather complicated in this
algebro-geometric context and we postpone its presentation for the next
section using a different approach.

Before concluding this section note 
that the $SO(3) \times SO(3)$ model, which arises as 
$\l_1 \rightarrow \l_2$, corresponds in this context to the
curve $w^2 =(\l_3-\l_2)v(v-1)^3(2v-1)$, which according to the
Riemann--Hurwitz relation has genus 1 as required; letting
$w \rightarrow w(v-1)$, we see that the cubic form 
$w^2 = (\l_3 -\l_2)v(v-1)(2v-1)$ results in this case.
Also, the $SO(5)$ model arises as $\l_2 \rightarrow \l_3$ and
it corresponds in this context to the curve
$w^2 =  (\l_2-\l_1)v^3(v-1)(2v-1)$, which again has genus 1, as required;
it transforms, in turn, into the cubic form $w^2 = (\l_2-\l_1)v(v-1)(2v-1)$
under the transformation $w \rightarrow wv$. Last, the $SO(4) \times SO(2)$
model is described by $w^2 = (\l_2 -\l_1) v(v-1)(2v-1)^2$ as
$\l_1 \rightarrow \l_3$. This has genus 0 and it can be brought into
a manifest quadratic form $w^2 = (\l_2-\l_1)v(v-1)$ using the transformation
$w \rightarrow w(2v-1)$. 
However, the bi-rational transformation for $y$ is appearing singular now, 
and the same is also true for the fully symmetric $SO(6)$ model; 
notice that for both of them the original form of the curve is not 
irreducible.  
Hence, we assume that the $SO(3) \times SO(2)$ model has $\l_1$, $\l_2$,
$\l_3$ all different from each other (in particular $\l_1 \neq \l_3$).
In any event, all previous models with genus 0 and 1 arise as special 
cases of $SO(3) \times SO(2)$ apart from the 
$SO(2) \times SO(2) \times SO(2)$ and the $SO(4)$ models 
that have already been described.

\section{Examples}
\setcounter{equation}{0}

The five- as well as the ten-dimensional forms of our solutions 
preserve four-dimensional 
Poincar\'e invariance $ISO(1,3)$ along the three-brane, but for general 
values of the constants $b_i$, they have no other continuous isometries.
In order to obtain some continuous group of isometries we have to choose 
some of the $b_i$'s equal. By means of \eqn{hja2} the corresponding scalars 
$\b_i$ are also equal to one another.
In this section we work out explicitly the 
expression for the metric and the scalar fields for some cases of 
particular interest using the ten-dimensional
geometric frame where $F$ is more 
conveniently regarded as a coordinate instead of using $z$. 
We will present the models with isometry groups $SO(3)\times 
SO(2)$ and its limiting cases $SO(3)\times SO(3)$ and $SO(5)$, as well
as the cases with isometry groups $SO(2)\times SO(2)\times SO(2)$, $SO(4)$
and their limiting model $SO(4)\times SO(2)$.
They describe all solutions with genus $\leq 2$ from the point of view of the
previous section. 
The examples are ordered by starting from the more general configurations 
and then specializing to models with higher symmetry.

The variable $z$ is more natural to use for addressing the uniformization
problem of the algebraic curves underlying in our solutions. In here,
we adapt our presentation to the ten-dimensional type-IIB supergravity 
description for two reasons: first as an alternative method for 
constructing explicit forms of our supersymmetric configurations,
and second for providing a higher dimensional view point for the
compactification to five space-time dimensions, and naturally for 
questions regarding the AdS/CFT correspondence. To avoid confusion 
note that certain choices of the moduli $b_i$ made in the sequel 
differ slightly from those made in the previous section, 
but this should cause no problem.

\subsection{Solutions with $SO(3)\times SO(2)$ symmetry}

In this case it is convenient to use a basis for the unit vectors 
$\hat x_i$ that define the five-sphere in such a way that it is in one to one
correspondence with the decomposition of 
the vector representation $\bf 6$ of $SO(6)$ with respect to the subgroup
$SO(3)\times SO(2)$, as ${\bf 6}\to ({\bf 3},{\bf 1})    
\oplus ({\bf 1},{\bf 2}) \oplus ({\bf 1},{\bf 1})$. 
Hence, we choose 
\ba 
\hat x_1 & =&  \cos\th \cos\psi\ ,\qq \pmatrix{\hat x_2\cr \hat x_3}\ = \  
\cos\th \sin\psi \pmatrix{\cos\varphi_1\cr\sin\varphi_1} \ ,
\nonumber\\
\pmatrix{\hat x_4 \cr \hat x_5}&= &  
\sin\th \sin\om\ \pmatrix{\cos\varphi_2 \cr\sin\varphi_2}\ ,\qq
\hat x_6 \ =\  \sin\th \cos\om\ .
\label{jwoi1}
\ea
It is also convenient to choose the constants $b_i$ as follows 
\be
b_1=b_2=b_3=0\ ,\qq b_4=b_5=-l_1^2\ ,\qq b_6=-l_2^2\ ,
\label{fdj1}
\ee   
where $l_1$ and $l_2$ are real constants, thus ordering now the moduli 
$b_i$ in an increasing order according to \eqn{hord1}.
We moreover adopt the change of variable $F=r^2$ with $r\ge 0$, 
which is legitimate as $b_{\rm max}=0$.
Then, the corresponding ten-dimensional metric takes the form 
\ba
ds^2& =& H^{-1/2} \eta_{\m\n} dx^\m dx^\n + H^{1/2}{\D \ov \D_1 \D_2}\ dr^2
\nonumber\\
&& +\  r^2 H^{1/2} \Big[ \big(\sin^2\th +\cos^2\th(\D_1 \sin^2\om + 
\D_2 \cos^2\om)\big) d\th^2 
\nonumber\\
&&+\ \cos^2\th d\Om_2^2  
 + \sin^2\th (\D_1 \cos^2\om + \D_2 \sin^2\om) 
d\om^2 + \sin^2\th \sin^2\om d\varphi_2^2 
\label{mee1}\\
&& +\ 2 \cos\th \sin\th \cos\om \sin\om (\D_1- \D_2) d\th d\om \Big]\ ,
\nonumber
\ea
where the various functions appearing in it are
\ba
\D_1 & =&  1+ {l_1^2\ov r^2}\ ,\qq \D_2 \ =\  1+ {l_2^2\ov r^2}\ ,
\nonumber\\
\D & =&  \D_1 \D_2 \cos^2\th + \sin^2\th (\D_1 \cos^2\om + \D_2 \sin^2\om)\ ,
\label{dh21}\\
H & = & 1 + {R^4 \D_2^{1/2} \ov r^4 \D}\ ,
\nonumber 
\ea 
and $d \Om_2^2$ is the two-sphere metric
\be
d\Om_2^2 = d\psi^2 + \sin^2\psi d\varphi_1^2\ .
\label{d2j1}
\ee

In terms of five-dimensional gauged supergravity, 
the five-dimensional metric \eqn{fh3} is described by the form 
\be
ds^2 = {\D_1^{1/3} \D_2^{1/6} r^2\ov R^2} \ \eta_{\m\n} dx^\m dx^\n 
+ {R^2\ov r^2 \D_1^{2/3} \D_2^{1/3}} \ dr^2 
\label{kao2}
\ee
and the expressions for the scalars \eqn{hja2} become
\ba
&&e^{2\b_1} = e^{2\b_2}= e^{2\b_3}=\D_1^{1/3} \D_2^{1/6}\ ,
\nonumber\\
&& e^{2\b_4} = e^{2\b_5}= \D_1^{-2/3} \D_2^{1/6}\ ,
\label{esc1}\\
&& e^{2\b_6}= \D_1^{1/3} \D_2^{-5/6}\ .
\nonumber
\ea

The metric \eqn{mee1} has a singularity at $r=0$ where the harmonic 
function $H$ diverges.
However, this is not a point-like singularity as it occurs for all possible
values of the angular variables $\th$, $\om$ and $\varphi_2$. 
Hence, \eqn{mee1} may be interpreted as representing the distribution of 
a large number of D3-branes inside a solid three-dimensional ellipsoid
defined by the equation
\be 
{y_4^2+y_5^2\ov l_1^2 } + {y_6^2\ov l_2^2} = 1 \ ,
\label{ell1}
\ee 
and the three-dimensional hyper-plane $y_1=y_2=y_3=0$.
We note that, by analytic continuation on the $l_i$'s we may obtain
brane distributions other than \eqn{ell1}, but we will not elaborate 
more on this point.

\subsection{Solutions with $SO(3)\times SO(3)$ symmetry}

In this case we may obtain the metric by just setting $l_1=l_2\equiv l$ in 
\eqn{mee1}, since then the symmetry is enhanced, from $SO(3)\times SO(2)$ to
$SO(3)\times SO(3)$. The metric becomes
\ba
ds^2 & =&  H^{-1/2} \eta_{\m\n} dx^\m dx^\n + H^{1/2} {r^2 + \l^2\cos^2\th\ov
r^2+ l^2}\ dr^2 
\nonumber\\
&& +\ H^{1/2} \left[ (r^2+l^2\cos^2 \th)d\th^2 
+ r^2\cos^2\th d\Om^2_2 + 
(r^2+l^2) \sin^2\th d\tilde\Om^2_2 \right]\ ,
\label{fe2}
\ea
where the harmonic function $H$ that follows from the corresponding 
expression in \eqn{dh21} is
\be
H = 1 + {R^4\ov r (r^2+l^2\cos^2\th)(r^2+l^2)^{1/2}}
\label{dfj3}
\ee
and the two different line elements for the two-dimensional sphere 
appearing in \eqn{fe2} are  
\be 
d\Om_2^2 = d\psi^2 + \sin^2\psi d\varphi_1^2\ ,\qq
d\tilde \Om_2^2 = d\om^2 + \sin^2\om d\varphi_2^2\ .
\label{d2d1}
\ee
The field theory limit form of the metric \eqn{fe2} (with the 1 omitted in 
\eqn{dfj3}) has appeared before in \cite{FGPW2}.

In the description in terms of gauged supergravity, 
the five-dimensional metric \eqn{fh3} takes the form 
\be
ds^2 = {r (r^2+l^2)^{1/2}\ov R^2} \ \eta_{\m\n} dx^\m dx^\n 
+ {R^2\ov r^2 + l^2 } \ dr^2 
\label{kao3}
\ee
and the expressions for the scalars \eqn{esc1} simplify to 
\ba
&& e^{2\b_1}= e^{2\b_2}= e^{2\b_3}= \left(1+{l^2\ov r^2}\right)^{1/2}\ ,
\nonumber\\
&& e^{2\b_4}= e^{2\b_5}= e^{2\b_6}= \left(1+{l^2\ov r^2}\right)^{-1/2}\ .
\label{esc51}
\ea
Note that the five-dimensional metric, as well as 
the corresponding scalar fields, take the equivalent form \eqn{a33} and
\eqn{b33}, respectively, when written in terms of the variable $z$.

Specializing \eqn{ell1} to the case at hand with $l_1=l_2=l$, we deduce 
that the metric \eqn{fe2} represents the distribution of a large number 
of D3-branes inside the solid three-dimensional ball
\be 
y_4^2+y_5^2+ y_6^2= l^2 \ ,
\label{spph1}
\ee 
in the three-dimensional hyper-plane defined by $y_1=y_2=y_3=0$.

\subsection{Solutions with $SO(5)$ symmetry}

In this case we may obtain the metric by just setting $l_1=0$ (and also
redefining $l_2\equiv l$) in 
\eqn{mee1}, since then the symmetry is enhanced from $SO(3)\times SO(2)$ to
$SO(5)$.
However, in order to present a metric with manifest $SO(5)$ symmetry, the basis
\eqn{jwoi1} is not appropriate.
A convenient basis for the unit vectors 
$\hat x_i$ should be such that it is in one to one correspondence with the 
decomposition of the vector representation $\bf 6$ of $SO(6)$, 
with respect to $SO(5)$, 
as ${\bf 6}\to {\bf 5} \oplus {\bf 1}$. Hence we choose
\ba
\pmatrix{\hat x_1\cr \hat x_2} & = & \cos\th \sin\psi \pmatrix{\cos\varphi_1\cr
\sin\varphi_1} \ ,
\nonumber\\
\pmatrix{\hat x_3\cr \hat x_4} & = & \cos\th \cos\psi \sin\om
\pmatrix{\cos\varphi_2\cr
\sin\varphi_2}\ ,
\nonumber \\
\hat x_5 & =&  \cos\th \cos\psi \cos\om \ ,
\nonumber \\
\hat x_6 & =&  \sin\th \ .
\label{jdj3}
\ea
The metric becomes
\ba
ds^2 & =&  H^{-1/2} \eta_{\m\n} dx^\m dx^\n + H^{1/2} 
{r^2 + l^2\cos^2\th\ov r^2+ l^2}\ dr^2 
\nonumber\\
&&  +\ H^{1/2} \left[(r^2+l^2\cos^2 \th)d\th^2+r^2\cos^2\th d\Om^2_4\right]\ ,
\label{fe24}
\ea
where the harmonic function is
\be
H = 1 + {R^4 (r^2+l^2)^{1/2}\ov r^3 (r^2+l^2\cos^2\th)}\ ,
\label{df2}
\ee
and the line element for the four-sphere is defined as
\be
d\Om_4^2= d\psi^2 + \sin^2\psi d\varphi_1^2 + \cos^2\psi(d\om^2 
+\sin^2\om d\varphi_2^2)\ .
\label{jkg2}
\ee
The field theory limit form of the metric \eqn{fe24} (with the 1 omitted in 
\eqn{df2}) has also appeared before in \cite{FGPW2}.

The five-dimensional metric \eqn{fh3} takes the form 
\be
ds^2 = {r^{5/3} (r^2+l^2)^{1/6}\ov R^2} \ \eta_{\m\n}dx^\m dx^\n + 
{R^2\ov r^{4/3} (r^2+l^2)^{1/3}}\ dr^2\ ,
\label{jfk4}
\ee
whereas the expressions for the scalars \eqn{esc1} become
\ba
&& e^{2\b_1}= e^{2\b_2}= e^{2\b_3}=e^{2\b_4}= e^{2\b_5}=
 \left(1+{l^2\ov r^2}\right)^{1/6}\ ,
\nonumber\\
&& e^{2\b_6}= \left(1+{l^2\ov r^2}\right)^{-5/6}\ .
\label{esc17}
\ea

The singularity of the metric \eqn{fe24} for $r=0$ may be interpreted as 
due to the presence of D3-branes distributed along the $y_6$ axis.
This can be also obtained from \eqn{ell1} in the limit
$l_1\to 0$ (and $l_2\equiv l$). In this limit, $y_4$ and $y_5$ are forced to 
be zero and therefore imposing \eqn{ell1} leads to $y_6=l$. 
Hence, the distribution of D3-branes is taken over a segment of length $l$.

\subsection{Solutions with $SO(2)\times SO(2)\times SO(2)$ symmetry}

In this case it is convenient to use a basis for the unit vectors 
$\hat x_i$ that define the five-sphere in such a way that it corresponds to
the decomposition of the vector representation 
$\bf 6$ of $SO(6)$ with respect to the full Cartan subgroup
$SO(2)\times SO(2)\times SO(2)$, as ${\bf 6}\to ({\bf 2},{\bf 1},{\bf 1})    
\oplus ({\bf 1},{\bf 2},{\bf 1}) \oplus ({\bf 1},{\bf 1},{\bf 2})$. 
Hence, we choose 
\ba 
\pmatrix{\hat x_1\cr \hat x_2} &= &  
\sin\th \pmatrix{\cos\varphi_1\cr\sin\varphi_1} \ ,
\nonumber\\
\pmatrix{\hat x_3 \cr \hat x_4}&= &  
\cos\th \sin\psi \pmatrix{\cos\varphi_2\cr\sin\varphi_2} \ ,
\label{jwoi}\\
\pmatrix{\hat x_5 \cr \hat x_6}& = & 
 \cos\th \cos\psi \pmatrix{\cos\varphi_3 \cr\sin\varphi_3}\ .
\nonumber
\ea
We also make the choice
\be
b_1=b_2 \equiv  a_1^2\ ,\qq b_3=b_4 \equiv  a_2^2\ ,\qq b_5=b_6 \equiv  
a_3^2\ ,
\label{fh1}
\ee
where $a_i$, $i=1,2,3$ are some real constants.

Using the change of variable $F=r^2$ (with $r\ge a_1$),
the metric is written as 
\ba
&& ds^2 = H^{-1/2} \eta_{\m\n} dx^\m dx^\n + H^{1/2}{\D r^4\ov f}\ dr^2
\nonumber\\
&&+ r^2 H^{1/2} 
\Bigg(\Delta_1 d\theta^2 +\Delta_2 \cos^2\theta d\psi^2 +
2 {a_2^2-a_3^2\over r^2}\cos\theta\sin\theta\cos\psi\sin\psi d\theta d\psi
\nonumber\\
&&+ (1-{a_1^2\over r^2})\sin^2\theta d\varphi_1^2 +
(1-{a_2^2\over r^2})
\cos^2\theta \sin^2\psi d\varphi_2^2 +
(1-{a_3^2\over r^2})\cos^2\theta\cos^2\psi d\varphi_3^2 \Bigg) 
\label{dsiib}
\ea
where the various functions are defined as
\ba
H & = & 1 + {R^4\ov r^4 \D}\ ,
\nonumber\\
f & = & (r^2-a_1^2)(r^2-a_2^2)(r^2-a_3^2)\ ,
\nonumber\\
\Delta &=& 1 -{a_1^2\over r^2} \cos^2\theta -{a_2^2\over r^2}
(\sin^2\theta\sin^2\psi +\cos^2\psi )
- {a_3^2\over r^2}(\sin^2\theta\cos^2\psi +\sin^2\psi )
\nonumber \\
&+& {a_2^2a_3^2\over r^4}\sin^2\theta +{a_1^2 a_3^2\over r^4}
\cos^2\theta\sin^2\psi +{a_1^2a_2^2\over r^4}\cos^2\theta\cos^2\psi\ ,
\label{d12}\\
\Delta_1
 &=& 1-{a_1^2\over r^2}\cos^2\theta -
{a_2^2\over r^2}\sin^2\theta\sin^2\psi -
{a_3^2\over r^2}\sin^2\theta\cos^2\psi\ ,
\nonumber\\
\Delta_2 &=& 1-{a_2^2\over r^2}\cos^2\psi -{a_3^2\over r^2}\sin^2\psi\ .
\nonumber 
\ea
The metric \eqn{dsiib}, together with the defining relations
\eqn{d12}, corresponds to the 
supersymmetric limit of the most general non-extremal rotating D3-brane 
solution \cite{RS1}. Using this interpretation, it turns out that 
$a_1,a_2$ and $a_3$ correspond to the three rotational
parameters of the solution, after a suitable Euclidean continuation.
We also note that the metric \eqn{dsiib} corresponds to the extremal limit 
of the three-charge black hole solution found in \cite{BCS}, in ansaetze
for solutions to $N=8$, $D=5$ gauged supergravity preserving an $U(1)^3$
subgroup of $SO(6)$ \cite{Cetall}.

The five-dimensional metric \eqn{aans} takes the form 
\be
ds^2 = {\prod_{i=1}^3 (r^2-a_i^2)^{1/3}\ov R^2}\ 
\eta_{\m\n} dx^\m dx^\n + {R^2 r^2\ov \prod_{i=1}^3 (r^2-a_i^2)^{2/3}}\ dr^2\ ,
\label{hh3h}
\ee
whereas the expressions for the scalar fields are 
\ba
&& e^{2 \b_1} = e^{2 \b_2} = (r^2-a_1^2)^{-2/3}  (r^2-a_2^2)^{1/3}
(r^2-a_3^2)^{1/3}\ ,
\nonumber\\
&& e^{2 \b_3} = e^{2 \b_3} = (r^2-a_1^2)^{1/3} (r^2-a_2^2)^{-2/3}
(r^2-a_3^2)^{1/3}\ ,
\label{bbb}\\
&& e^{2 \b_5} = e^{2 \b_6} = (r^2-a_1^2)^{1/3} (r^2-a_2^2)^{1/3}
(r^2-a_3^2)^{-2/3}\ ,
\nonumber
\ea

The relationship to elliptic functions is made explicit by first using 
the definition \eqn{e222}, which is rewritten here 
in terms of three parameters $a_i$ as
\be
e_i = a_i^2 -{1\ov 3} (a_1^2 + a_2^2 + a_3^2) \ ,\qq i=1,2,3\ .
\label{wkp1}
\ee
Then, the differential equation \eqn{jds1} has as solution the 
Weierstrass elliptic function ${\cal P}$ 
\be
F(z)=  {\cal P}(z/R^2)\ ,
\label{soo1}
\ee
which is the same as \eqn{f222} after ignoring the irrelevant additive
constant. The invariants of the curve that define the Weierstrass elliptic 
function ${\cal P}$ are 
\be
g_2= -4(e_1 e_2 + e_2 e_3 + e_3 e_1)\ ,\qq g_3 = 4 e_1 e_2 e_3\ .
\label{jr2}
\ee
Since the Weierstrass function ${\cal P}$ 
is double periodic with half-periods $\om_1$ and $\om_2$ given by 
\be
\om_1 = {{ K}(k)\ov \sqrt{e_1-e_3}} \ ,\qq
\om_2 = {i { K}(k')\ov \sqrt{e_1-e_3}}\ ,
\label{dawh1}
\ee
where ${ K}$ is the complete elliptic integral 
of the first kind with modulus $k$ and complementary modulus $k'$,
we arrive at the following identification in terms of the rotational parameters
\ba
&& k^2={e_2-e_3\ov e_1-e_3}={a_2^2-a_3^2\ov a_1^2-a_3^2}\ ,
\nonumber \\
&& k'^2= 1-k^2= {e_1-e_2\ov e_1-e_3}= {a_1^2-a_2^2\ov a_1^2-a_3^2}\ .
\label{dh1}
\ea
Finally, after changing variable 
\be
r = {\sqrt{a_1^2 - a_3^2} \ov {\rm sn} u}\ ,
\qq u \equiv {\sqrt{a_1^2 - a_3^2}\ov R^2}\ z
\ ,
\label{ej1}
\ee
where ${\rm sn} u$ is the Jacobi function,
the metric \eqn{hh3h} assumes the conformally flat form \eqn{aans} with 
\be
e^A = {\sqrt{a_1^2-a_3^2}\ov R}\ {{\rm cn}^{1/3}u\ {\rm dn}^{1/3}u\ov 
{\rm sn} u} = {1\ov R} \left({\cal P}^\prime(z/R^2)\ov 2\right)^{1/3}\ .
\label{fj3}
\ee
The last equality describes precisely the result found in \eqn{s222}
using the algebro-geometric method of uniformization. 
Also, in terms of the variable $z$, the scalar fields \eqn{bbb} 
coincide with \eqn{s222}.

\subsection{Solutions with $SO(4)\times SO(2)$ symmetry}

These solutions can be obtained by letting $e_2=e_3$ (equivalently $a_2=a_3$) 
into the various 
expressions of the previous subsection. In this limit, 
by taking into account the change of radial variable 
as $r^2\to r^2 + a_2^2$, the metric \eqn{dsiib} becomes 
\ba
&& ds^2  = H^{-1/2} \eta_{\m\n} dx^\m dx^\n
+ H^{1/2} {r^2-r_0^2\cos^2\th\ov r^2-r_0^2}\ dr^2 
\nonumber \\
&& +  H^{1/2} \left( (r^2-r_0^2\cos^2\th) 
\Big({dr^2\ov r^2-r_0^2}+ d\th^2 \Big)
+ (r^2-r_0^2) \sin^2\th d\varphi_1^2 
+r^2 \cos^2\th d\Omega_3^2\right)
\label{ruu1}
\ea 
where $r_0^2 \equiv a_1^2-a_2^2$ and the harmonic function is 
\ba
H & =&  1 + {R^4\ov r^2 (r^2-r_0^2 \cos^2\th)} 
\nonumber\\
& = & 1+ {2 R^4\ov \sqrt{(r_6^2-r_0^2)^2+4 r_0^2 r_2^2} 
\left( r_6^2 + r_0^2 + \sqrt{(r_6^2-r_0^2)^2+4 r_0^2 r_2^2} \right) } \ ,
\label{dj32}
\ea
where $r_6^2=y_1^2+\dots + y_6^2$ and $r_2^2=y_1^2+y_2^2$. 
In the second line of \eqn{dj32} we have written for completeness 
the harmonic function $H$
in terms of the Cartesian coordinates by explicitly substituting 
the function $F$ as a solution of the condition \eqn{jk4}. 
The result agrees with what was obtained
previously in \cite{KLT,sfe1}. The three-sphere line element that appears
in \eqn{ruu1} is given by 
\be
d\Om_3^2 = d\psi^2 + \sin^2\psi d\varphi_2^2 +  \cos^2\psi d\varphi_3^2\ .
\label{lpj2}
\ee

The five-dimensional metric \eqn{aans} takes the form \cite{BS2}
\be
ds^2 = {r^{4/3} (r^2-r_0^2)^{1/3}\ov R^2}\ 
\eta_{\m\n} dx^\m dx^\n + {R^2 \ov r^{2/3} (r^2-r_0^2)^{2/3}}\ dr^2\ ,
\label{h3h}
\ee
whereas the expressions for the scalar fields are given by
\ba
&& e^{2 \b_1} = e^{2 \b_2} = \left(1-{r_0^2\ov r^2}\right)^{-2/3} \ ,
\nonumber\\
&& e^{2 \b_3} = e^{2 \b_4} = e^{2 \b_5} = e^{2 \b_6}=
\left(1-{r_0^2\ov r^2}\right)^{1/3}\ .
\label{bb4}
\ea
Assuming that $r_0^2> 0$, we find that the metric \eqn{ruu1} 
has a singularity at
$r=r_0$ and $\th=0$. This is not a point-like singularity as it occurs for 
general values of $\psi,\varphi_2$ and $\varphi_3$.
It describes the situation where the horizon of the 
non-extremal metric coincides with the singularity as one approaches 
the extremal limit.
The singularity of the metric \eqn{ruu1} can be interpreted as
arising from the presence of D3-branes distributed on a spherical 
shell \cite{KLT,sfe1}
defined in the $y_1=y_2=0$ hyper-plane by the equation
\be
y_3^2+y_4^2+y_5^2+y_6^2=r_0^2\ .
\label{ppo1}
\ee

In the case that $e_1=e_2$ (equivalently $a_2=a_1$) it turns out that the 
previous results apply with $r_0^2= a_3^2-a_1^2<0$. 
It is then appropriate to define a new positive parameter by just 
letting $r_0^2\to -r_0^2$. Then, the singularity of the metric \eqn{ruu1}
occurs at $r=0$ and may be interpreted as coming from  
the presence of D3-branes distributed over a disc \cite{KLT,sfe1},
whose boundary is defined in the $y_3=y_4=y_5=y_6=0$ hyper-plane by
the circle 
\be
y_1^2+y_2^2=r_0^2\ .
\label{fjh4}
\ee

It is instructive to recover the metric of 
five-dimensional gauged supergravity
corresponding to our solution as a limiting case of 
\eqn{fj3}, in analogy with the limiting description of the ten-dimensional 
metric \eqn{ruu1}. To comment on this, let us first consider the
limiting case $e_3=e_2$ (or equivalently 
$a_3=a_2$), where the modulus $k\simeq 0$ and the elliptic curve degenerates
along the $a$-cycle. 
Then, using the well known properties of the Jacobi functions 
${\rm cn u}\simeq \cos u$, ${\rm sn u}\simeq \sin u$ and ${\rm dn u}\simeq 1$,
that are valid for $k\simeq 0$, we obtain from \eqn{fj3}
that the conformal factor in the 
corresponding five-dimensional metric \eqn{aans} is given by
\be
e^{2A} = {r_0^2\ov R^2} {\cos^{2/3} (r_0 z/R^2)\ov \sin^2(r_0 z/R^2)}\ , 
\label{dj9}
\ee
where the variable $u$ in \eqn{ej1} becomes 
$u=r_0/R^2 z$, with $r_0^2=a_1^2-a_2^2$, when $a_3=a_2$. 
Another limiting case arises when $e_2=e_1$ (or equivalently 
$a_2=a_1$), in which the complementary modulus $k'\simeq 0$ and the 
elliptic curve degenerates along the $b$-cycle. 
Then, using the properties of the Jacobi functions 
for $k'\simeq 0$ we have
${\rm cn u}\simeq 1/\cosh u$, 
${\rm sn u}\simeq \tanh u$ and ${\rm dn u}\simeq 1/\cosh u$.
>From \eqn{fj3} we obtain that the conformal factor of the 
corresponding five-dimensional metric \eqn{aans} becomes
\be
e^{2A} = {r^2_0\ov R^2} {\cosh^{2/3} (r_0 z/R^2)\ov \sinh^2(r_0 z/R^2)}\ , 
\label{dj89}
\ee
where we have used the fact that 
the variable $u$ in \eqn{ej1} becomes $u=r_0/R^2 z$, with $r_0$ now 
defined as $r_0^2= a_1^2-a_3^2$, when $a_2=a_1$. 
We note that the conformal factors appearing in \eqn{dj9} and \eqn{dj89}
are the same as those found before in \cite{BS2}.
They are also precisely the same factors as those appearing 
in \eqn{a42} and \eqn{a42a} by appropriate identification of
the parameters and after 
reinstating the scale factor $R$ into the equations.

\subsection{Solutions with $SO(4)$ symmetry}

In this case we choose four of the constants $b_i$ equal to each other
as follows
\be 
b_1=-l_1^2\ ,\qq b_2= - l_2^2 \ ,\qq b_3=b_4=b_5=b_6 = 0\ .
\label{jwef3}
\ee
Using the basis \eqn{jwoi} for the $\hat x_i$'s we find that the metric takes
the form
\ba
 ds^2 & = & H^{-1/2} \eta_{\m\n} dx^\m dx^\n +\ H^{1/2}{\D \ov \D_1 \D_2}\ dr^2
\nonumber\\
&& +\ r^2 H^{1/2} \Big[ \big(\sin^2\th +\cos^2\th(\D_1 \cos^2\varphi_1 + 
\D_2 \sin^2\varphi_1)\big) d\th^2 
\nonumber\\
&&+\ \cos^2\th d\Om_3^2  
 + \sin^2\th (\D_1 \sin^2\varphi_1 + \D_2 \cos^2\varphi_1) d\varphi_1 
\label{merhj}\\
&&+\ 2 \cos\th \sin\th \cos\varphi_1 \sin\varphi_1 
(\D_2- \D_1) d\th d\varphi_1\Big]\ ,
\nonumber
\ea
where 
\ba 
\D_1 & = & 1+{l_1^2\ov r^2} \ ,\qq \D_2 \ = \ 1+{l_2^2\ov r^2}\ ,
\nonumber\\
\D & = & \D_1 \D_2 \cos^2\th +\sin^2\th 
(\D_1 \sin^2\varphi_1+\D_2 \cos^2\varphi_1)\ ,
\label{jkasf1}\\
H & = & 1+ { R^4 \D_1^{1/2} \D_2^{1/2}\ov r^4 \D}\ .
\nonumber
\ea

The five-dimensional gauged supergravity metric \eqn{aans} becomes
\be
ds^2 = {r^2 \D_1^{1/6} \D_2^{1/6}\ov R^2} \ \eta_{\m\n} dx^\m dx^\n + 
{R^2\ov r^2 \D_1^{1/3} \D_2^{1/3}} \ dr^2 \ ,
\label{jfh6}
\ee
and the scalar fields are given by
\ba
&& e^{2 \b_1} = \D_1^{-5/6} \D_2^{1/6}\ ,
\nonumber\\
&& e^{2 \b_2} = \D_1^{1/6} \D_2^{-5/6}\ ,
\label{bb47}\\
&& e^{2 \b_3} = e^{2 \b_4} = e^{2 \b_5} = e^{2 \b_6}= \D_1^{1/6} \D_2^{1/6}\ .
\nonumber
\ea

The metric \eqn{merhj} has a singularity at $r=0$, where the harmonic function 
$H$ in \eqn{jkasf1} blows up. It can be interpreted as being
due to a continuous 
distribution of D3-branes in the ellipsoidal disc defined by 
\be
{y_1^2\ov l_1^2} + {y_2^2\ov l_2^2} = 1\ ,
\label{fj33}
\ee 
lying in the $y_3=y_4=y_5=y_6=0$ hyper-plane. 
Note also that in the case when $l_1=l_2$, the symmetry is 
enhanced from $SO(4)$
to $SO(4)\times SO(2)$. Then, the expressions for the metric \eqn{merhj} 
and the scalars fields \eqn{bb47} coincide with those found in \eqn{ruu1} and 
\eqn{bb4} respectively using the identification $r_0^2= -l_1^2=-l_2^2$.
Also, when one of the $l_i$'s becomes zero, the symmetry is enhanced from
$SO(4)$ to $SO(5)$ and by a suitable change of coordinates one recovers the 
results of subsection 4.3. 

\section{Spectrum for scalar and spin-two fields} 
\setcounter{equation}{0}

In this section we investigate the problem of solving the differential 
equations for the massless scalar field as well as for the
graviton fluctuations in our 
general five-dimensional background metric \eqn{aans}. We 
formulate the problem in terms of an equivalent Schr\"odinger equation 
in a potential that depends on the particular background. 
Later in this section we will discuss explicitly some cases of particular
interest and determine the exact form of the corresponding potentials.

\subsection{Generalities}

We begin with the massless scalar field equation 
$\Box_5 \Phi =0$ in the background geometry \eqn{aans}.
In the context of the AdS/CFT correspondence, 
the solutions and eigenvalues 
of this equation have been associated with the spectrum of the operator
${\rm Tr} F^2$ \cite{GKT,Witten,FFZ}.
On the other hand, the fluctuations of the graviton polarized in the 
directions parallel to the brane are associated 
with the energy momentum tensor $T_{\m\n}$ on the gauge theory side 
\cite{GKT,Witten,FFZ}.
A priori, one expects that 
the spectra of the two operators ${\rm Tr} F^2$ and $T_{\m\n}$ are 
different.
However, as was shown in \cite{BS2}, when graviton fluctuations on a
three-brane embedded in a five-dimensional metric as in \eqn{aans}
are considered, 
the two spectra and the corresponding eigenfunctions coincide.
In particular, in order to study the graviton fluctuations,
the Minkowski metric $\eta_{\m\n}$ along the three-brane 
is replaced in \eqn{aans}
by $\eta_{\m\n} + h_{\m\n}$ and then the 
equations of motion \eqn{eqs32} are linearized in $h_{\m\n}$. 
Reparametrization invariance allows to gauge-fix five functions.
In the gauge $\del_\m h^{\m}{}_\n=h^\m{}_\m =0$,
where indices are raised and lowered using $\eta_{\m\n}$ and its inverse,
the graviton fluctuations 
obey the equation $\Box_5 h_{\m\n}=0$, which is the same equation as that 
for a scalar field \cite{BS2} (the same 
observation has been made in a slightly different context in \cite{CM-BMT}).
Hence, the spectra for the operators ${\rm Tr} F^2$ and $T_{\m\n}$ 
indeed coincide. In what follows, $\Phi$ will denote either a massless
scalar field or any component of the graviton tensor field.

We proceed further by making the following ansatz for the solution 
\be 
\Phi(x,z) = \exp(i k\cdot x) \phi(z)\ ,
\label{hd22}
\ee
which represents 
plane waves propagating along the three-brane with an amplitude function that 
is $z$-dependent.
The mass-square is defined as $M^2 = - k\cdot k$. 
Using the expression for the metric in \eqn{aans}, we find that the equation
for $\phi(z)$ is 
\be
\phi^{\prime\prime} + 3 A^\prime \phi^\prime + M^2 \phi = 0 \ .
\label{acc2}
\ee
This equation can be cast into a Schr\"odinger equation for a wavefunction
$\Psi(z)$ defined as $\Psi= e^{3 A/2} \phi$. We find 
\be
-\Psi^{\prime\prime} + V \Psi = M^2 \Psi \ ,
\label{ss2}
\ee
with potential given by 
\be
V  =  {9\ov 4} {A^{\prime}}^2 + {3\ov 2} A^{\prime\prime}\ .
\label{jwd}
\ee
So far our discussion is quite general and applies to all solutions of the
system of equations \eqn{eqs32}. When the solutions are supersymmetric we
may use \eqn{ai1}, \eqn{hja1} and \eqn{jds1} to find 
alternative forms for the potential, namely
\ba 
V & = & {e^{2 A}\ov 16 R^2}\left[ 3 \Big(\sum_{i=1}^6 e^{2 \b_i}\Big)^2 
- 8 \sum_{i=1}^6 e^{4 \b_i} \right] 
\nonumber \\
& = & {f^{1/2}\ov 16 R^4} \left[3 \Big(\sum_{i=1}^6 {1\ov F-b_i}\Big)^2 
-8 \sum_{i=1}^6 {1\ov (F-b_i)^2}\right]\ .
\label{hf8}
\ea
This expression for the potential depends, of course, on the variable 
$z$ through the function $F(z)$. Even without having knowledge of 
the explicit $z$-dependence
of the potential, we may deduce some general properties about the spectrum 
in the various cases of interest. 
Further details will be worked out in the following subsection
using the results of section 4.

In general, $F$ takes values between the maximum 
of the constants $b_i$ (which according to the ordering made in \eqn{hord1}
is taken to be $b_1$) and $+\infty$. 
When $F\to \infty$, the five-dimensional space approaches
$AdS_5$ and the potential becomes
\be
V\simeq {15 F\ov 4 R^4}\ , \qq {\rm as} \quad F\to \infty\ ,
\label{hjf4}
\ee
and hence it is unbounded from above. 
Let us now consider the behaviour of the potential close to the other end,
namely when $F\to b_1$. Consider the general case 
where $b_1$ appears $n$ times, as
in the corresponding discussion made at the end of subsection 3.1.
Using \eqn{hf8} we find that the potential 
behaves (including the subscript $n$ 
to distinguish the various cases) as
\be
V_n\simeq {f_0^{1/2}\ov 16 R^4}\ n (3 n -8) (F-b_1)^{{n\ov 2}-2}\ ,\qq
{\rm as} \quad F\to b_1\ ,
\label{hjfg3}
\ee
with $f_0$ being a constant given, as before, by $f_0=\prod_i(b_1-b_i)$.
Hence, for the value $n=6$, corresponding to $AdS_5$ the potential goes 
to zero and the spectrum is continuous. The same is true for the value $n=5$
corresponding to the $SO(5)$ symmetric model. For the case $n=4$ the potential 
approaches a constant value with the metric given by \eqn{merhj}.
Using the definitions \eqn{jwef3}, the general expression for the 
the minimum value of the potential is in this case
\be
V_{4,{\rm min}}={l_1 l_2\ov R^4 } \ . 
\label{hf56}
\ee
Therefore, although the spectrum is continuous,
it does not start from zero, but there is a mass gap whose value squared
is given by the minimum of the potential in \eqn{hf56}. 
For the $SO(4) \times SO(2)$ model, where the metric is given by 
\eqn{ruu1} with $r_0^2<0$, the existence of a mass gap 
has already been proven in \cite{FGPW2,BS1}.
For $n=3$ the potential goes to $+\infty$ as $F\to b_1$
and therefore the spectrum is not continuous but discrete.
Quite generally we may show, using simple scaling arguments, that the
typical unit of mass square is $f_0^{1\ov 6-n}/R^4$. Hence, for $n=3$, 
we expect that $M^2$ will be quantized in units of $f_0^{1/3}/R^4$.
For $n=2$ the potential goes to $-\infty$ and there is the danger that 
it is unbounded from below. Nevertheless, at least for the
$SO(4) \times SO(2)$ symmetric model
with metric given by \eqn{ruu1} with $r_0^2>0$,
it was shown before that $M^2$ is discrete and positive 
\cite{FGPW2,BS1}.

\subsection{Examples of potentials}

\subsubsection{  \underline{$SO(6)$}}

Let us consider first the massless scalar equation for the most symmetric
case, namely when the background is given by the $AdS_5$ metric itself.
In this case we have $e^{A} =R/z$ and the potential becomes
\be
V(z) = {15\ov 4 z^2} \ , \qq 0 \le z < \infty\ ,
\label{hj11}
\ee
which obviously has a continuous spectrum for $M^2$. 
The corresponding Schr\"odinger equation can be transformed into a 
Bessel equation and the result for the amplitude of the fluctuations \eqn{hd22}
is given by 
\be
\phi_M = (MR)^{-3/2} (Mz)^2 J_2(M z)\ ,
\label{bell2}
\ee
where $J_2$ is the Bessel function of index 2, which is regular at the origin 
$z=0$. The arbitrary overall constant is chosen so that the Dirac-type
normalization condition is satisfied
\be
\int_0^\infty dz e^{3A} \phi_M \phi_{M'} = \d(M-M')\ .
\label{noormm}
\ee
The measure factor $e^{3A}$ in the integrand of the equation above
is such that the Schr\"odinger wave function $\Psi = e^{3 A/2} \phi$ obeys
a normalization condition similar to \eqn{noormm}, but with measure 1.

\subsubsection{ \underline{$ SO(4)\times SO(2)$}}

Consider now the first non-trivial case with metric given by 
\eqn{ruu1} with $r_0^2>0$. 
The spectrum for massless scalar and graviton 
fluctuations has already been analyzed in \cite{FGPW2,BS1} and \cite{BS2},
respectively. We include this case here not only for completeness,
but also because we will make connections with Calogero-type models later. 
The explicit form for the potential turns out to be 
\be
V(z) = {r_0^2\ov R^4} \left(-1- {1\ov 4 \cos^2(r_0z/R^2) }
+{15\ov 4 \sin^2(r_0 z/R^2)} \right) \ ,\qq 0\le z \leq {\pi R^2\ov 2 r_0}\ ,
\label{j1d}
\ee
and clearly possesses the features we have discussed at the begining 
of this section.
In fact \eqn{j1d} belongs to a family of potentials called  
P\"oschl--Teller potential of type I in the literature
of elementary quantum mechanics.
The solution for the massless scalar or graviton fluctuations is given by 
\eqn{hd22} with the quantized amplitude modes being given by \cite{BS1}
\be
\phi_n = \sqrt{(2 n+3) r_0\ov 8 R^2} (1-x)^2 P_{n}^{(2,0)}(x)\ ,\quad
x=1-2 {r_0^2\ov r^2}= \cos(2 r_0 z/R^2)\ ,\quad n=0,1,\dots \ ,
\label{jdkl3}
\ee
where in general $P^{(\a,\b)}_n$ denote the classical Jacobi polynomials. 
Note that the arbitrary overall 
constant in \eqn{jdkl3} has been chosen so that the $\phi_n$'s are 
normalized to 1 with measure $e^{3A}$, similar to \eqn{noormm}, where 
$A$ is now given by \eqn{dj9}.
The associated mass spectrum is
\be 
M^2_n= {4r_0^2\ov R^4} (n+1)(n+2)\ ,\qq n=0,1,\dots \ .
\label{jdf9}
\ee

Let us now turn to the case of the metric \eqn{ruu1} with $r_0^2<0$ and
replace $r_0^2$ by $-r_0^2$. Then, the potential takes the form 
(also given in \cite{BS2})
\be
V(z) = {r_0^2\ov R^4} \left(1+ {1\ov 4 \cosh^2(r_0 z/R^2) }
+{15\ov 4 \sinh^2(r_0 z/R^2)} \right) \ ,\qq 0\le z < \infty \ ,
\label{jd}
\ee
which is the so called P\"oschl--Teller potential of type II. 
Note that this potential is related to the one appearing in 
\eqn{j1d} by analytic
continuation $r_0\to i r_0$, as expected.
This potential approaches the value $r_0^2/R^4$ as $z\to \infty$ 
and therefore its spectrum is continuous, but with
a mass gap given by \cite{FGPW2,BS1}
\be
M^2_{\rm gap} = {r_0^2\ov R^4} \ . 
\label{hgef3}
\ee
In this case the solution for the massless scalar and the
graviton fluctuations is given by \eqn{hd22} with amplitude 
\ba
\phi_q \sim  x^{(q-1)/2} F_q(x)  - x^{-(q+1)/2} F_{-q}(x) \ , \quad 
0\leq x= {r^2\ov r^2+r_0^2}= {1\ov \cosh^2(r_0 z/R^2)} \leq 1\ .
\label{duw}
\ea
The constant $q$ and the function $F_q(x)$ are related 
to the mass $M$ via hypergeometric functions as
\be
F_q(x)  = F\Big({q-1\ov 2},{q-1\ov 2},1+q; x\Big)\ ,\qq
q = \sqrt{1-R^2 M^2} \ .
\label{gqe}
\ee
Note that the constant $q$ is purely imaginary
due to the mass gap of the model.

\subsubsection{\underline{$ SO(2)\times SO(2)\times SO(2)$}}

Since this case has not been discussed in the literature, we
will explain the derivation of the potential $V(z)$ in some detail.
Using the variable $u = z/R^2$, for simplicity, we find according
to the definition that the potential becomes
\ba
V(u) &=& {1 \over 16 R^4}{{\cal P}^{\prime}}^2(u)\Bigg(3\left(
{1 \over {\cal P}(u) -e_1} + {1 \over {\cal P}(u) - e_2} +
{1 \over {\cal P}(u) -e_3}\right)^2 \nonumber\\ 
& & -{4 \over ({\cal P}(u) -e_1)^2}
-{4 \over ({\cal P}(u) -e^2)^2} -{4 \over ({\cal P}(u) -e^3)^2}
\Bigg) ,
\ea
where $e_1 + e_2 + e_3 =0$ for the roots of the corresponding elliptic
curve. Then, using the addition theorem for the Weierstrass function we
have
\ba
& &{1 \over 4}\left({{{\cal P}^{\prime}}^2(u) \over ({\cal P}(u) -e_1)^2}
+ {{{\cal P}^{\prime}}^2(u) \over ({\cal P}(u) -e_2)^2} 
+ {{{\cal P}^{\prime}}^2(u) \over ({\cal P}(u) -e_3)^2} \right) =
\nonumber\\
& &3{\cal P}(u) +{\cal P}(u+\omega_1) +{\cal P}(u+\omega_2) 
+{\cal P}(u+\omega_1 + \omega_2) 
\ea
and so a straightforward calculation yields the final result
\be
V(z) = {1 \over 4R^4}\left(15{\cal P}\left({z \over R^2}\right) - 
{\cal P}\left({z \over R^2} + \omega_1\right) - 
{\cal P}\left({z \over R^2} + \omega_2\right)
-{\cal P}\left({z \over R^2} + \omega_1 + \omega_2\right) \right) 
\ee
with the dependence on $R$ appearing now explicitly. 

It is easy to see how
the degeneration of the curve leads to the rational potential of
the $SO(4) \times SO(2)$ model. Recall that for $e_1 \neq e_2 = e_3$,
i.e. for elliptic modulus $k =0$, the $a$-cycle of the Riemann surface
shrinks to zero size and the Weierstrass function simplifies to
\be
{\cal P}(u) =-{3g_3 \over 2g_2} +{9g_3 \over 2g_2} 
{1 \over {\rm sin}^2\left(u \sqrt{{9g_3 \over 2g_2}}\right)} ~.
\ee
In this limiting case we have $9g_3 / 2g_2 = a_1^2 - a_2^2$, whereas
$a_2 = a_3$, using the rotational parameters of our ten-dimensional 
solution. Since this combination equals to $r_0^2$, we obtain the 
trigonometric function ${\rm sin}(r_0z/R^2)$. In this limit we
also have $\omega_1 = \pi /2r_0$ and $\omega_2 = i \infty$, and so
${\cal P}(u + \omega_1)$ will involve the function
${\rm cos}(r_0z/R^2)$, while the terms originating from 
${\cal P}(u +\omega_2)$ and
${\cal P}(u +\omega_1 +\omega_2)$ contribute only to the constant.
In this fashion we recover the potential of the $SO(4) \times SO(2)$
model. Its hyperbolic counterpart appears when $r_0^2 <0$ and so 
the new potential can be obtained by suitable analytic continuation.

Note finally that in the general case 
the potential becomes infinite at the Weierstrass
points 0, $\omega_1$, $\omega_2$, $\omega_1 + \omega_2$, 
because the Weierstrass function blows up at 0 modulo the periods;
put differently, some term of the potential becomes infinite at each 
one of these points. Unfortunately, we do not have complete grasp
of the spectrum for the Schr\"odinger equation in this potential.
We hope that its computation will be discussed elsewhere.

\subsubsection{\underline{$SO(3) \times SO(3)$}}

This case also leads to a new form for the potential that has not
been investigated before. Again, using for simplicity the parameter
$u=z/(2 R^2)$, since for $R=1$ the uniformizing parameter is $u=-z/2$,
and the minus sign plays no r\^ole in $V$,
 we have for our solution (with general $R$)
\be
V(u) = {3 \over 256 R^4}{{{\cal P}^{\prime}}^2(u) \over {\cal P}^2
(u)} \left( \left({{\cal P}(u) -b_1 +b_2 \over {\cal P}(u) +b_1-b_2}
\right)^2 +\left({{\cal P}(u) + b_1 -b_2 \over {\cal P}(u) 
-b_1 +b_2}\right)^2 + 18 \right) .
\ee
This follows by substitution of our algebro-geometric solution into
the defining relation of the potential, after reinstating the 
$R$-dependence. The elliptic curve has presently $g_3 = 0$ and so
$e_2 = 0$, $e_1 = - e_3$. By employing the identities, special to 
this surface,
\be
{\cal P}^{\prime}(u+\omega_1) = -2e_1^2 {{\cal P}^{\prime}(u) \over
({\cal P}(u) - e_1)^2} ~, ~~~~~ 
{\cal P}^{\prime}(u+\omega_2) = -2e_3^2 {{\cal P}^{\prime}(u) \over
({\cal P}(u) - e_3)^2} ~,
\ee
where $e_1^2 = e_3^2 = (b_1-b_2)^2$, we arrive after some calculation
at the final result for the potential
\be
V(z) = {3 (b_1 -b_2)^4 \over 4R^4} \left({1 \over 
{{\cal P}^{\prime}}^2\left({z \over 2 R^2} + \omega_1\right)} + {1 \over
{{\cal P}^{\prime}}^2\left({z \over 2 R^2} + \omega_2\right)} + {18 \over
{\cal P}^{\prime}\left({z\over 2 R^2} + \omega_1\right) 
{\cal P}^{\prime}\left({z \over 2 R^2} + \omega_2\right)} \right) .
\ee

Note that this potential also becomes infinite at the four Weierstrass
points, but its structural dependence on elliptic functions seems to be 
different from the previous example. 
However, making use of some further identities (special to the curves
with $g_3 =0$), it can be cast into a form proportional to
\be
{15 \over 4} {\cal P} \left({z \over 2R^2} \right) + {3 \over 4} {\cal P} 
\left({z \over 2R^2} + \omega_1 \right) + {15 \over 4} {\cal P}
\left({z \over 2R^2} 
+ \omega_1 + \omega_2 \right) + {3 \over 4} {\cal P} \left({z \over 2R^2} + 
\omega_2 \right) ~,  
\ee
which is invariant under shifts with respect to $\omega_1 + \omega_2$. 
We leave the computation of the
spectrum for the corresponding Schr\"odinger equation 
to future investigation, as before.  

For the other examples we have been unable to derive the form of the
potential in closed form, because there are no closed formulae for 
the solutions in terms of the variable $z$ that appears naturally
in the corresponding Schr\"odinger equation, or else in the 
algebro-geometric description of the various models as Riemann surfaces.
We close this section with some general remarks concerning the rational 
and elliptic
variations of Calogero-like models.

\subsection{Comments}

It is rather amusing that the Schr\"odinger problem one has to solve
in $z$ is of Calogero type. This is a characteristic feature of $AdS_5$ 
and possibly of more general $AdS$ spaces.
For the maximally symmetric $SO(6)$ model, in particular, the potential 
is precisely $V(z) = 1/z^2$ (up to an overall scale) \eqn{hj11}. 
It can be seen 
that the solutions of the less symmetric $SO(4) \times SO(2)$ model 
are also related to a Calogero problem, 
namely the three-body model in one dimension.  
Recall that the quantum states of the general three-body problem 
can be found by solving the time independent Schr\"odinger equation
\be
\left(- \sum_{i=1}^3 {{\partial}^2 \over \partial x_i^2} + 2g
\sum_{i, j = 1}^3 {1 \over (x_i - x_j)^2} + 6f \sum_{i, j, k = 1}^3
{1 \over (x_i + x_j -2 x_k)^2} - E\right) {\Psi}(x_1, x_2, x_3) = 0 \ ,
\ee
for ${i \neq j \neq k}$,
where the $x_i$'s describe the coordinates of the three particles and 
$g$, $f$ denote the strength of the two-body and three-body Calogero
interactions respectively. 
In fact, from a group theory point of view, this potential describes
couplings between the particles according to the root system of 
the simple Lie algebra $G_2$. Introducing the center of mass coordinates
\be
\sqrt{2} ~ r {\rm sin}\varphi= x_1 - x_2 ~, ~~~~ 
\sqrt{6} ~ r {\rm cos}\varphi = x_1 + x_2 -2x_3 ~, ~~~~
3R_{\rm cm} = x_1 + x_2 +x_3\ ,
\ee
one obtains, after moding out the $R_{\rm cm}$-dependence, 
a differential equation 
for the wavefunction ${\Psi}(r, \varphi)$. It can be separated, as usual,
into two independent equations for the radial and angular dependence
of the wavefunctions
\be
\left(- {{\partial}^2 \over \partial r^2} - {1 \over r}{\partial \over
\partial r} + {{\lambda}^2 \over r^2} - E\right) X(r) = 0 ~,
\ee
\be
\left(-{{\partial}^2 \over \partial {\varphi}^2} + {9g \over 
{\rm sin}^2 (3\varphi)} + {9f \over {\rm cos}^2 (3\varphi)} - {\lambda}^2
\right) \Phi(\varphi) = 0 ~,
\ee
where ${\lambda}^2$ is the separation constant, 
${\Psi}(r, \varphi) = X(r) \Phi (\varphi)$, and $E$ is now the energy in
the center of mass frame \cite{Ref1}. 
Of course, the Hamiltonian is Hermitian provided
that the coupling constants $g \geq -1/4$ and $f \geq -1/4$. For $f = 0$
only two-body interactions are present in the problem. 

The integrability of the classical 
Calogero model persists quantum mechanically 
and helps us to determine its spectrum and wave 
eigenfunctions. In particular, the differential
equation for the angular dependence is solved in general as follows,
\ba
&& {\Phi}_n(\varphi) = {\rm sin}^{{\mu}_1} (3\varphi)
{\rm cos}^{{\mu}_2} (3\varphi)
P_n^{{\mu}_1 -1/2 , {\mu}_2 -1/2}({\rm cos}(6\varphi))  \ ,
\nonumber \\
&& {\lambda}_n^2 = 9(2n + {\mu}_1 + {\mu}_2)^2 ~, \qq n = 0, 1, 2, \dots \ ,
\ea
where the ${\mu}_i$'s are introduced as
\be
g = {\mu}_1 ({\mu}_1 -1) ~, ~~~~~~ f = {\mu}_2({\mu}_2 -1) ~.
\ee
The angle $\varphi$ assumes the values between 0 and $\pi /6$; because
of its dependence on the Cartesian coordinates $x_i$, a particular value
of $\varphi$ gives a specific ordering of the three particles and hence
the problem can be divided into sectors depending on the range of
$\varphi$.
For a general overview of these issues, see for instance \cite{Ref2} and 
references therein.
Interestingly enough, the Schr\"odinger problem that arose in studying
the spectrum of quantum fluctuations for scalar and spin-two fields in 
the background of the $SO(4) \times SO(2)$ model of five-dimensional
gauged supergravity
fits precisely into the integrable class of such Calogero potentials with
${\mu}_1 = 5/2$ and ${\mu}_2 = 1/2$, which thus attains its minimum 
value required by hermiticity. Note, however, that presently 
${\mu}_1 \neq {\mu}_2$. To make exact contact with our problem for 
the $SO(4)\times SO(2)$ model,
first introduce the necessary rescaling with respect to $R$,
setting $3\varphi = r_0 z/R^2$, and then
conclude that the mass spectrum is given in general by
\be
M_n^2 = {r_0^2 \over R^4}\left({{\lambda}_n^2 \over 9} - 1\right) \ ,\qq
 n= 0, 1, 2, \dots ~,
\label{jf5}
\ee
as there is also a constant term which is present now 
that shifts the energy levels.
For the values ${\mu}_1 = 5/2$ and ${\mu}_2 = 1/2$ the spectrum \eqn{jf5}
coincides with that in \eqn{jdf9}.

On the other hand, the elliptic generalization 
of the $1 /z^2$ potential arose 
historically more than a century ago in connection with the problem of
finding ellipsoidal harmonics for the 3-dim Laplace
equation. When one deals with physical problems connected to ellipsoids,
like having sources with a general ellipsoidal distribution, 
the mathematical structure of spheres, cones and planes usually associated
to polar coordinates gets replaced by the structure of confocal quadrics.
Since the transformation from Cartesian coordinates is not singled valued,
elliptic functions are employed for its proper description.
Introducing uniformizing parameters associated with confocal coordinates,
it turns out that the solutions of Laplace's equation are obtained by
separation of variables, in which case one arrives at the Lam\'e
equation
\be
\left(-{d^2 \over dz^2} + n(n+1) {\cal P}(z) - E\right) \Psi(z) = 0 ~
\ee
for harmonics of degree $n$; $E$ is a separation constant that appears
in the mathematical analysis of the problem (for details see
\cite{Ref3}). It is interesting to note that this particular
Schr\"odinger problem was fully investigated much later in connection
with finite zone potentials, Riemann surfaces and the KdV hierarchies
(see for instance \cite{Ref4} and references therein), since the Weierstrass
function satisfies the time independent KdV equation.
It comes as no surprize that potentials consisting of Weierstrass 
functions also arise in our study, because the relevant configurations 
can be obtained from distributions of D3-branes on ellipsoids in 
ten dimensions, as it has already been noted in the geometrical setting
of our solutions. In this sense,
all potentials that occur in the supergravity models with genus $g>0$ 
should be considered as appropriate generalizations of the original 
derivation of Lam\'e's equation in a ten-dimensional IIB framework. 

Multi-particle systems with two-body interactions described by
${\cal P}(z_i - z_j)$ have also been studied extensively as integrable
systems \cite{Ref2}. 
However, the trigonometric identities used earlier for expressing
${\rm sin} 3\varphi$ and ${\rm cos} 3\varphi$ in terms of 
${\rm sin}\varphi$ and ${\rm cos}\varphi$ for the rational 
three-body Calogero model, thus arriving at a separation
of the angular $\varphi$ dependence, are not generalizable to 
elliptic functions. Hence, there is no analogous understanding
of the Schr\"odinger equation that determines the spectrum of scalar and
spin-two fields in the five-dimensional
background of our elliptic configurations
using many-body elliptic Calogero systems. To the best of our knowledge,
the specific quantum problems that arise here have not been investigated 
and pose a set of interesting questions for future work.

We mention for completeness that the only problem which has been studied
in detail among the class of potentials given by a sum of Weierstrass
functions concerns the Schr\"odinger equation with
\be
V(z) = 2\sum_{i=1}^n {\cal P}(z-z_i(t))
\ee
when $z_i(t)$ are moduli that evolve in time as elliptic Calogero particles
with two-body interactions only, namely
\be
{d^2 z_i(t) \over dt^2} = 4\sum_{i \neq j} {\cal P}^{\prime}(z_i - z_j) ~.
\ee
Such systems are naturally encountered in the description of elliptic
solutions of the KP equation, in analogy with the rational solutions of
the KP equation where the ordinary $1/z^2$ Calogero models make their
appearance (see for instance \cite{Ref5}). 
A static solution is easily obtained by considering
four such particles located at the corners of a parallelogram inside the
fundamental domain of elliptic functions described by the points
$z_1 = 0$, $z_2 = {\omega}_1$, 
$z_3 = {\omega}_2$ and $z_4 = {\omega}_1 + {\omega}_2$. In this case,
their
differences $z_i -z_j$ equal to half-periods (modulo the periods) 
for all $i \neq j$, so the derivative of the Weierstrass 
function vanishes there
and the elliptic Calogero equations are trivially satisfied.
Then, the potential for the Schr\"odinger equation becomes
\be
V(z) = 2\left({\cal P}(z) + {\cal P}(z+{\omega}_1) +
{\cal P}(z+{\omega}_2) +{\cal P}(z+{\omega}_1 + {\omega}_2)\right) , 
\ee
which by the way equals to $8 {\cal P}(2z)$ and reduces to the usual
Lam\'e equation with $n=1$ after rescaling $z$.
The generalization to potentials consisting of similar 
Weierstrass terms but
with more arbitrary relative coefficients, as in the $SO(2)\times 
SO(2)\times SO(2)$ model, or as in the
$SO(3)\times SO(3)$ model, remain
open for study and we hope to return elsewhere in view of their
relevance in five-dimensional gauged supergravity.

\section{Conclusions}

In this paper we have analyzed the conditions for having 
supersymmetric configurations in five-dimensional gauged supergravity
in the sector where only five scalar fields, associated
with the coset space $SL(6,\IR)/SO(6)$, are turned on apart from
the metric. These conditions were integrated using an 
ansatz for the conformal factor of the five-dimensional metric in terms of
a function $F(z)$, and the scalar fields were subsequently 
determined provided that a certain non-linear differential 
equation for the function $F(z)$ could be solved. This approach 
provides a natural algebro-geometric framework in which 
Riemann surfaces and their uniformization play a prominent
role. A key ingredient was the interpretation of the non-linear 
differential equation for $F(z)$ as a Schwatz--Christoffel 
transformation by extending the range of parameters to the complex
domain. In fact, the general solution depends on six real moduli,
which when they start coalescing lead to configurations with 
various symmetry groups. The case with maximal symmetry $SO(6)$
corresponds to the maximally supersymmetric solution of $AdS_5$
with all scalar fields set equal to zero. More generally, 
we have classified all such algebraic
curves according to their genus, and associated symmetry groups
(all being subgroups of $SO(6)$). 
We also made use of their uniformization 
for finding explicit forms of the supersymmetric states in terms
of elliptic functions. The calculations have been carried out in 
detail for the models of low genus, but they can be extended to all
other cases with higher genus (or else smaller symmetry groups).

There is an alternative description of our solutions in terms of 
type-IIB supergravity in ten dimensions, which is a natural place 
for discussing solutions of five-dimensional gauged supergravity via 
consistent truncations. This higher dimensional point of view
is also interesting for addressing various questions related to the
AdS/CFT correspondence and supersymmetric Yang-Mills theory in four
space-time dimensions. We found that the algebraic classes of our
five-dimensional configurations could be understood as representing the
gravitational field of a large number of D3-branes continuously
distributed on hypersurfaces embedded in the six-dimensional 
space that is
transverse to the branes. The geometry of these hypersurfaces is
closely related to the Riemann surfaces underlying in the 
algebro-geometric approach, as the distribution of D3-branes 
is taken to be in the interior of certain ellipsoids for the
corresponding elliptic solutions. Also, as more and more scalar 
fields are turned on, the geometry of the five-dimensional sphere that 
appears in the ten-dimensional description 
of our states (together with the remaining five dimensions 
which are asymptotic to $AdS_5$ space) becomes deformed and respects
less and less symmetry from the original $SO(6)$ symmetry group 
of the round $S^5$. In this geometrical approach, there is no need to
perform the uniformization of Riemann surfaces, as the metric is
formulated in another frame with $F(z)$ being the coordinate 
variable instead of $z$. 
The Schwarz--Christoffel transform describes precisely this particular 
change of coordinates, when it is restricted to real values.
Then, the calculation reduces to finding appropriate harmonic
functions that correspond to the continuous distribution of D3-branes.
In any event, both approaches are equivalent to each other and 
complement nicely the classification of the supersymmetric states
that has been considered.

Finally, we have examined the spectra of the massless scalar and 
graviton fields on these backgrounds and found that they can be
determined by a Schr\"odinger equation in one dimension, which
is $z$, with a potential that depends on the conformal factor of the
five-dimensional metric. It is rather curious that all these potentials are
essentially 
of Calogero type. In the fully symmetric $SO(6)$ model, whose solution
represents $AdS_5$, the potential is $1/z^2$, which is a characteristic
feature of Calogero systems. 
For other models with less symmetry, the potential turns 
out to be either a rational form of Calogero interactions
or elliptic generalizations thereof depending on each case. 
Such generalized potentials were not investigated in the
literature before and there are many questions that are left open
concerning their integrability properties and the exact determination 
of the spectrum. We think that supersymmetric quantum
mechanics could help to make progress in this direction.

It will be also interesting to consider in future study the precise
characterization of all these states 
in connection  with the representation
theory of the complete supersymmetry algebra. Shrinking cycles that
lower the genus of our algebraic curves and lead to enhancement of
the symmetry group of the various 
models should have an interesting interpretation
in more traditional terms, using the representations of supersymmetry
and the associated multiplets. Moreover, the extention of our
techniques to other theories of gauged supergravity, in particular
in higher dimensions, seems possible and we hope to return to all
these elsewhere.

\bigskip\bigskip

\centerline{ \bf Acknowledgements}

One of the authors (I.B.) wishes to thank CERN/TH for hospitality
and support during the course of this work. He is also grateful
to the organizers of the summer institute at Ecole Normale
Superieure for their kind invitation to present a preliminary version
of these results and for stimulating conversations.


\begin{thebibliography}{3}

\bibitem{cremmer}
E. Cremmer, {\it Supergravities in 5 dimensions}, edited by
S.W. Hawking and M. Rocek,
Proceedings of the Nuffield Gravity Workshop, 
Cambridge, June 16-July 12, Cambridge Univiversity Press 1981.

\bibitem{PPN}
M. Pernici, K. Pilch and P. van Nieuwenhuizen, Nucl. Phys. {\bf B259}
(1985) 460.

\bibitem{GRW}
M. Gunaydin, L.J. Romans and N.P. Warner, Phys. Lett. {\bf 154B}
(1985) 268 \break 
and Nucl. Phys. {\bf B272} (1986) 598.

\bibitem{CJ} 
E. Cremmer and B. Julia, Nucl. Phys. {\bf B159} (1979) 141.

\bibitem{WN1}
B. de Wit and H. Nicolai, Nucl. Phys. {\bf B208} (1982) 323.


\bibitem{Maldacena}
J. Maldacena,
Adv. Theor. Math. Phys. {\bf 2} (1998) 231, {\tt hep-th/9711200}.

\bibitem{Witten}
E. Witten,
Adv. Theor. Math. Phys. {\bf 2} (1998) 253, {\tt hep-th/9802150}.


\bibitem{GKP}
S.S. Gubser, I.R. Klebanov and A.M. Polyakov,
Phys. Lett. {\bf B428} (1998) 105, {\tt hep-th/9802109}.


\bibitem{MW}
J.A. Minahan and N.P. Warner,
JHEP {\bf 06} (1998) 005, {\tt hep-th/9805104}.

\bibitem{KW}
I.R. Klebanov and E. Witten, Nucl. Phys. {\bf B556} (1999) 89,
{\tt hep-th/9905104}.


\bibitem{KLT}
P. Kraus, F. Larsen and S.P. Trivedi,
JHEP {\bf 03} (1999) 003, {\tt hep-th/9811120}.

\bibitem{sfe1}
K. Sfetsos,
JHEP {\bf 01} (1999) 015, {\tt hep-th/9811167}.


\bibitem{FGPW2}
D.Z. Freedman, S.S. Gubser, K. Pilch and N.P. Warner,
{\it Continuous distributions of D3-branes and gauged supergravity},
{\tt hep-th/9906194}.


\bibitem{BS1}
A. Brandhuber and K. Sfetsos,
{\it Wilson loops from multicentre and rotating branes, 
mass gaps and phase  structure in gauge theories}, to appear in 
Adv. Theor. Math. Phys.,
{\tt hep-th/9906201}.


\bibitem{CR-GR}
I. Chepelev and R. Roiban, Phys. Lett. {\bf B462} (1999) 74, 
{\tt hep-th/9906224};\break
S.B. Giddings and 
S.F. Ross, Phys. Rev. {\bf D61} (2000) 024036,
{\tt hep-th/9907204}.

\bibitem{GPPZ1}
L. Girardello, M. Petrini, M. Porrati and A. Zaffaroni,
JHEP {\bf 12} (1998) 022, {\tt hep-th/9810126}.

\bibitem{DZ}
J. Distler and F. Zamora,
Adv. Theor. Math. Phys. {\bf 2} (1999) 1405, {\tt hep-th/9810206}.

\bibitem{KPW}
A. Khavaev, K. Pilch and N.P. Warner,
{\it New vacua of gauged N = 8 supergravity in five dimensions},
{\tt hep-th/9812035}.


\bibitem{FGPW1}
D.Z. Freedman, S.S. Gubser, K. Pilch and N.P. Warner,
{\it Renormalization group flows from holography supersymmetry and 
a c-theorem}, {\tt hep-th/9904017}.

\bibitem{GW}
V.A.~Rubakov and M.E.~Shaposhnikov,
Phys. Lett. {\bf 125B} (1983) 136;
G.W.~Gibbons and D.L.~Wiltshire, Nucl. Phys. {\bf B287} (1987) 717 and
references therein.

\bibitem{rasu}
L.~Randall and R.~Sundrum, Phys. Rev. Lett. {\bf 83} (1999) 3370,
and Phys.Rev. Lett. {\bf 83} (1999) 4690.

\bibitem{BS2}
A. Brandhuber and K. Sfetsos, JHEP {\bf 10} (1999) 013,
{\tt hep-th/9908116}.







\bibitem{KS1}
A. Kehagias and K. Sfetsos,
Phys. Lett. {\bf B454} (1999) 270, {\tt hep-th/9902125};\break
S.S. Gubser,
{\it Dilaton-driven confinement}, 
{\tt hep-th/9902155};
L. Girardello, M. Petrini, M. Porrati and A. Zaffaroni,
JHEP {\bf 05} (1999) 026, {\tt hep-th/9903026}.


\bibitem{witnik}
B. de Wit and H. Nicolai, Nucl. Phys. {\bf B281} (1987) 211;
B. de Wit, H. Nicolai and N.P. Warner, Nucl. Phys. {\bf B255} (1984) 29.

\bibitem{KaKou}
R. Kallosh and J. Kumar,
Phys. Rev. {\bf D56} (1997) 4934, {\tt hep-th/9704189}.


\bibitem{RS1}
J.G. Russo and K. Sfetsos,
Adv. Theor. Math. Phys. {\bf 3} (1999) 131, {\tt hep-th/9901056}.

\bibitem{BCS}
K. Behrndt, M. Cvetic and W.A. Sabra,
Nucl. Phys. {\bf B553} (1999) 317, {\tt hep-th/9810227}.

\bibitem{Cetall}
M.~Cvetic,
M.J. Duff, P. Hoxha, J.T. Liu, H. Lu,
J.X. Lu, R. Martinez-Acosta, C.N. Pope, H. Sati and T.A. Tran,
{\it Embedding AdS black holes in ten and eleven dimensions},
{\tt hep-th/9903214}.


\bibitem{GKT}
S.S. Gubser, I.R. Klebanov and A.A. Tseytlin,
Nucl. Phys. {\bf B499} (1997) 217, {\tt hep-th/9703040}.

\bibitem{FFZ}
S. Ferrara, C. Fronsdal and A. Zaffaroni,
Nucl. Phys. {\bf B532} (1998) 153,\hfill\break {\tt hep-th/9802203}.

\bibitem{CM-BMT}
N.R. Constable and R.C. Myers, JHEP {\bf 10} (199) 037,
{\tt hep-th/9908175};
R.C. Brower, S.D. Mathur and C. Tan,
{\it Discrete spectrum of the graviton in the AdS(5) black hole background},
{\tt hep-th/9908196}.

\bibitem{Ref1} J. Wolfes, J. Math. Phys. {\bf 15} (1974) 1420; 
F. Calogero and C. Marchioro, J. Math. Phys. {\bf 15} 
(1974) 1425.

\bibitem{Ref2} M. Olshanetsky and A. Perelomov, Phys. Rep. {\bf 94}
(1983) 313.

\bibitem{Ref3} E. Whittaker and G. Watson, ``A Course of Modern Analysis" 
(fourth edition), Cambridge University Press (1927).

\bibitem{Ref4} B. Dubrovin and S. Novikov, Sov. Phys. JETP 
{\bf 40} (1974) 1058; A. Its and V. Matveev, Theor. Math. 
Phys. {\bf 23} (1975) 343; 
B. Dubrovin, Russ. Math. Surv. {\bf 36} (1981) 11.

\bibitem{Ref5} I. Krichever, Funct. Anal. Appl. {\bf 12} (1978) 76;
ibid. {\bf 14} (1980) 45; J. Sov. Math. {\bf 28} 
(1985) 51.



\end{thebibliography}
\end{document}